\documentclass[amsfonts,amsmath,prd,preprint,nofootinbib]{revtex4}
\newcommand{\beq}{\begin{equation}}
\newcommand{\eeq}{\end{equation}}

\newcommand{\bsp}{\begin{split}}

\usepackage{epsfig,bbm,cancel,ulem}
\usepackage[breaklinks=true]{hyperref}
\usepackage{xcolor}
\usepackage{wasysym}
\usepackage[mathscr]{eucal}
\usepackage{multirow}

\begin{document}

\title{Geodesic of nonlinear electrodynamics and stable photon orbits}

\author{A S. Habibina}
\email{a.sayyidina@sci.ui.ac.id}
\author{H.~S.~Ramadhan}
\email{hramad@sci.ui.ac.id}
\affiliation{Departemen Fisika, FMIPA, Universitas Indonesia, Depok, 16424, Indonesia. }
\def\changenote#1{\footnote{\bf #1}}

\begin{abstract}

We study the geodesics of charged black holes in polynomial Maxwell lagrangians, a subclass models within the nonlinear electrodynamics (NLED). Specifically, we consider black holes in Kruglov, power-law, and Ayon-Beato-Garcia models. Our exploration on the corresponding null bound states reveals that photon can orbit the extremal black holes in stable radii outside the corresponding horizon, contrary to the case of Reissner-Nordstrom (RN) black holes. The reason behind this is the well-known theorem that photon in NLED background propagates along its own {\it effective} geometry. This nonlinearity is able to shift the local minimum of the effective potential away from its corresponding outer horizon. For the null scattering states we obtain corrections to the weak deflection angle off the black holes. We rule out the power-law model to be physical since its deflection angle does not reduce to the Schwarzschild in the limit of vanishing charge.
\end{abstract}

\maketitle
\thispagestyle{empty}
\setcounter{page}{1}

\section{Introduction}

One of the many intriguing properties of black hole (BH) is the notion of {\it photon sphere}, the path upon which null rays can orbit in constant radius. The recent profound observation of black hole is made possible by producing the ring-like image of photon sphere around the supermassive BH~\cite{Akiyama:2019cqa}. This discovery relies on the rather realistic rotating BH whose (circular as well as spherical) photon orbits have extensively been investigated, for example in~\cite{Teo, Bugden:2018uya} and the references therein. It is nevertheless also of high interest to study photon sphere in static cases. The Schwarzschild is known to have (unstable) null orbit at $r=3M$ in natural unit. The RN black hole possesses two photon spheres $r_{ps}^{\pm}={3M\over2}\left(1\pm\sqrt{1-{8\over9}\left({Q\over M}\right)^2}\right)$, only one of which ($r_+$) can be observed since it lies outside its outer horizon. This physical orbit is the local maximum of the corresponding effective null potential; thus also unstable. As pointed out in~\cite{Pradhan:2010ws, Khoo:2016xqv} the extremal RN can have stable photon orbit exactly {\it on} its (extreme) horizon, $r_{EH}$. The meaning ``{\it stable}" here, we argue, is still controversial since any small perturbation around $r_{EH}$ shall collapse the photon inside the horizon. Naturally, one would expect that this condition can be cured in the NLED case.

Nonlinear electrodynamics (NLED) is not new in modern physics. Mie in 1912, and later Born and Infeld in 1934, proposed that electron is a nonsingular solution of field theory with finite electromagnetic energy~\cite{ Chernitskii:2005ih, Born:1934gh}. With the development of quantum electrodynamics (QED) this classical nonlinear field theories were later abandoned. Ironically, it is precisely the success of QED that resurrects the recent interest in NLED. Recent photon-photon scattering experimental results strongly indicate that in the vacuum electrodynamics might be nonlinear~\cite{Bamber:1999zt, Tommasini:2008lrs, Pike:2014wha, Gaete:2014nda}. Euler and Heisenberg predicted that vacuum magnetic birefringence must occur in QED~\cite{Heisenberg:1935qt}. This phenomenon is absent in Maxwell and Born-Infeld (BI) electrodynamics, but can be present in other NLED. The bound for the birefringence's magnitude is provided by the BMV and and PVLAS experiments~\cite{Battesti:2012hf, Cadene:2013bva, DellaValle:2014xoa}, and improvements are still being sought. It is then no wonder that in a recent decade there is abundant proposals for NLED. Kruglov proposed a generalization of BI electrodynamics as a model of fractional electrodynamics~\cite{Kruglov:2009he, Kruglov:2016uzf} and a few nonlinear electrodynamics model with trigonometric terms \cite{Kruglov:2015fcd,Kruglov:2019wjv}. Euler-Heisenberg electrodynamics which features second order of Maxwell electrodynamics was revisited in \cite{Kim:2011fy, Kruglov:2017ymn}. Logarithmic electrodynamics was investigated in~\cite{Gaete:2013dta, Kruglov:2014iqa}, while exponential electrodynamics along with its phenomenology were studied in~\cite{Sheykhi:2017yqt, Kruglov:2017fck, Sheykhi:2016aoi}. The first black hole solutions coupled with nonlinear charge was discussed by Hoffmann and Infeld and by Peres~\cite{Hoffmann:1937noa, Peres:1961zz}. They presented exact solutions of Einstein-Born-Infeld (EBI) theory. Today several exact solutions of black holes charged with NLED sources, both in general relativity (GR) as well as in modified gravity, have extensively been explored (see, for example,~\cite{Fernando:2003tz, Hassaine:2008pw, Hassaine:2007py, Bronnikov:2000vy, Kruglov:2017xmb, Breton:2001yk, Kruglov:2017mpj, Hendi:2014bba, Hendi:2014kha, Jana:2015cha} and references therein). The cosmological effect of a form of NLED with one parameter was examined in \cite{Kruglov:2015fbl,Samanta:2018cgn}. Some of the most studied models of NLED are the conformally invariant power Maxwell electrodynamics which was analyzed as higher-dimensional black holes in \cite{Hendi:2010bk,Sheykhi:2012zz} and the Ayon-Beato-Garcia electrodynamics  which was developed using Hamiltonian formulation to construct an electrically-charged Bardeen's regular black hole solution in \cite{AyonBeato:1998ub} and the magnetically-charged one on \cite{AyonBeato:2000zs}, where it was resurfaced in recent studies \cite{Ghaffarnejad:2016dlw,Rodrigues:2018bdc}. 

It is well-known that photon behaves differently in the NLED ambient compared to the linear Maxwell electrodynamics. They do not propagate along the background geometry's null geodesic. Rather, they follow the null geodesic of its effective geometry~\cite{Novello:1999pg}. This behavior gives shed on the photon sphere study on charged BH. Curiously, research on this topic is rather rare\footnote{The relations between photon spheres in Einstein-BI gravity with its phase transitions are studied in~\cite{Xu:2019yub, Li:2019dai} and the references therein}. It is therefore of our interest to study this phenomenology in the vast literature of NLED models. In this work we shall investigate the null geodesic of several NLED models in the framework of GR. For simplicity, in this preliminary work we shall focus on the polynomial Maxwell-type lagrangians. That is, we consider three models: the Kruglov, the power-law Maxwell, and the Ayon-Beato-Garcia NLED models.

This work is organized as follows. In Section~\ref{overview} we give a brief overview of general NLED model. Sections~\ref{GBI_GR}-\ref{ABG_GR} are devoted to investigating the three different NLED models. In each we study their timelike and null geodesics, as well as the weak deflection angle of light. We completely produce the photon orbit values for each model. Our conclusion is summarized in Section~\ref{conclude}.

\section{Overview of NLED}
\label{overview}

In general, all NLED models can be expressed as a functional of Maxwell's Lagrangian, $\mathcal{L}=\mathcal{L}[\mathcal{F}]$, where\footnote{Throughout this work we shall not deal with $\mathcal{G}\equiv\frac{1}{4}F_{\mu\nu} \tilde{F}^{\mu\nu}$. This can be done by setting its constant parameter to be zero.} $\mathcal{F}\equiv\frac{1}{4}F_{\mu\nu} F^{\mu\nu}$. By correspondence principle, in the low-energy /weak-coupling limit they all should reduce to Maxwell, $\mathcal{L}=-\mathcal{F}$. 

Any nonlinearization extension of an established theory must obey causality and unitary principles. In the context of electrodynamics they can be formulated as the following constraints~\cite{Shabad:2011hf, Kruglov:2017fuj}:
\begin{equation}
\mathcal{L_F}\leq0,\ \ \ \mathcal{L_{FF}}\geq0,\ \ \ \mathcal{L_F}+2\mathcal{F}\mathcal{L_{FF}}\leq0,
\end{equation}
where $\mathcal{L_F}\equiv\partial\mathcal{L}/\partial\mathcal{F}$ and $\mathcal{L_{FF}}\equiv\partial^2\mathcal{L}/\partial\mathcal{F}^2$.

The field equation is given by its corresponding Euler-Lagrange,
\begin{equation}
\label{NLEDeq}
\nabla_{\mu}\left(\mathcal{L_F}F^{\mu\nu}\right)=0.
\end{equation}
Alternatively, one can define, by means of Legendre transformation, the corresponding ``Hamiltonian"~\cite{Salazar:1987ap, AyonBeato:1998ub}
\begin{equation}
\mathcal{H}\equiv2\mathcal{L_F F-L}.
\end{equation}
The Lagrangian, in turn, can be written as
\begin{equation}
\mathcal{L}=2\mathcal{H_P P-H},
\end{equation}
where $\mathcal{P}\equiv{1\over4}P_{\mu\nu}P^{\mu\nu}$ and $P^{\mu\nu}\equiv\mathcal{L_F}F^{\mu\nu}$. It is easy to see that $\mathcal{H}=\mathcal{H}[\mathcal{P}]$. The field equation is then given by
\begin{equation}
\nabla_{\mu}P^{\mu\nu}=0.
\end{equation}

In any case, the field equations can be shown to be 
\begin{equation}
\label{DH}
\nabla\cdot{\bf D}=0,\ \ \ {\partial{\bf D}\over\partial t}=\nabla\times{\bf H},
\end{equation}
with ${\bf D}\equiv\partial\mathcal{L}/\partial{\bf E}$ the electric displacement field and ${\bf H}\equiv-\partial\mathcal{L}/\partial{\bf B}$ the magnetic field. The nonlinearity implies the relation between ${\bf E}$ and ${\bf D}$ as well as ${\bf B}$ and ${\bf H}$ not linear. In general, ${\bf D}={\bf D}\left({\bf E}, {\bf B}\right)$ and ${\bf H}={\bf H}\left({\bf E}, {\bf B}\right)$. One interesting phenomenological interpretation is that NLED describes the electromagnetic wave propagation in nonlinear media.

\section{Generalized Born-Infeld}
\label{GBI_GR}

This model was proposed by Kruglov to generalize the BI Electrodynamics \cite{Kruglov:2016uzf}, 
\begin{equation}
\label{gbi_nled}
\mathcal{L} _K= \frac{1}{\beta} \bigg[ 1- \bigg( 1+ \frac{\beta F}{q}\bigg)^q\bigg].
\end{equation}
Here $\beta$ is a parameter with dimension of $[L]^4$ and $q$ is an arbitrary dimensionless parameter. For $q=1$ the model reduces to Maxwell, while $q=1/2$ gives us BI electrodynamics. In flat spacetime the field equation~\ref{NLEDeq} yields 
\begin{equation}
\partial_{\mu}\left(\Gamma^{q-1}F^{\mu\nu}\right)=0,\ \ \ \ \ \  \Gamma\equiv1+{\beta\mathcal{F}\over q}.
\end{equation}
This equation is equivalent to~\ref{DH} with the following identifications
\begin{equation}
{\bf D}=\varepsilon{\bf E},\ \ \ \ {\bf H}=\mu^{-1}{\bf B},
\end{equation}
where $\varepsilon=\mu^{-1}\equiv\Gamma^{q-1}$. For electric point-charge source the displacement field ${\bf D}(r)$ is singular at the origin, but the electric field ${\bf E}(r)$ is not. It is regular at the core with finite value given by ${\bf E}(0)=\sqrt{2q\over\beta}$.

The monopole black hole can be obtained from the Einstein-Kruglov model~\cite{Kruglov:2017mpj} 
\begin{equation}
\label{action_GR}
S = \int d^4 x \sqrt{-g}\bigg[\frac{R}{2\kappa^2} + \mathcal{L}_K\bigg],
\end{equation}
where $\kappa^2\equiv8\pi G$. The ansatz employed here is magnetic monopole and spherical symmetry~\cite{Bronnikov:2000vy},
\begin{equation}
\label{A_mu_ms}
A_{t} = A_{r} = A_{\theta} =0, \,\,\,\,\,\,\,\,  \,\,\,\,\,\,\,\,  A_{\phi} = Q (1-\cos\theta),
\end{equation}
and
\begin{equation}
\label{flat_geo}
ds^2 = -f(r) dt^2 +f^{-1}(r) dr^2 + r^2 d\Omega^2.
\end{equation}

The solutions are~\cite{Kruglov:2017mpj}
\begin{equation}
F_{\theta\phi} = Q \sin \theta,
\end{equation}
and
\begin{eqnarray}
\label{gbi_f2r}
f(r) = 1 -\frac{2 M}{r} -\frac{\kappa ^2 r^2}{3 \beta} \bigg[\, _2F_1\left(-\frac{3}{4},-q;\frac{1}{4};-\frac{Q^2 \beta }{2 q r^4}\right)-1\bigg],
\end{eqnarray}
where $Q$ is the magnetic charge and $_2F_1\left(a,b;c;d\right)$ the hypergeometric function. It can be shown that in the limit of $\beta\rightarrow 0$ the solution reduces to the magnetic RN, while for $q\rightarrow1/2$ it reduces to the magnetic BI black holes~\cite{Fernando:2003tz, Breton:2001yk}.
\begin{figure}[!ht]
	\centering
	\begin{tabular}{cc}
		\includegraphics[height=6.3cm,keepaspectratio]{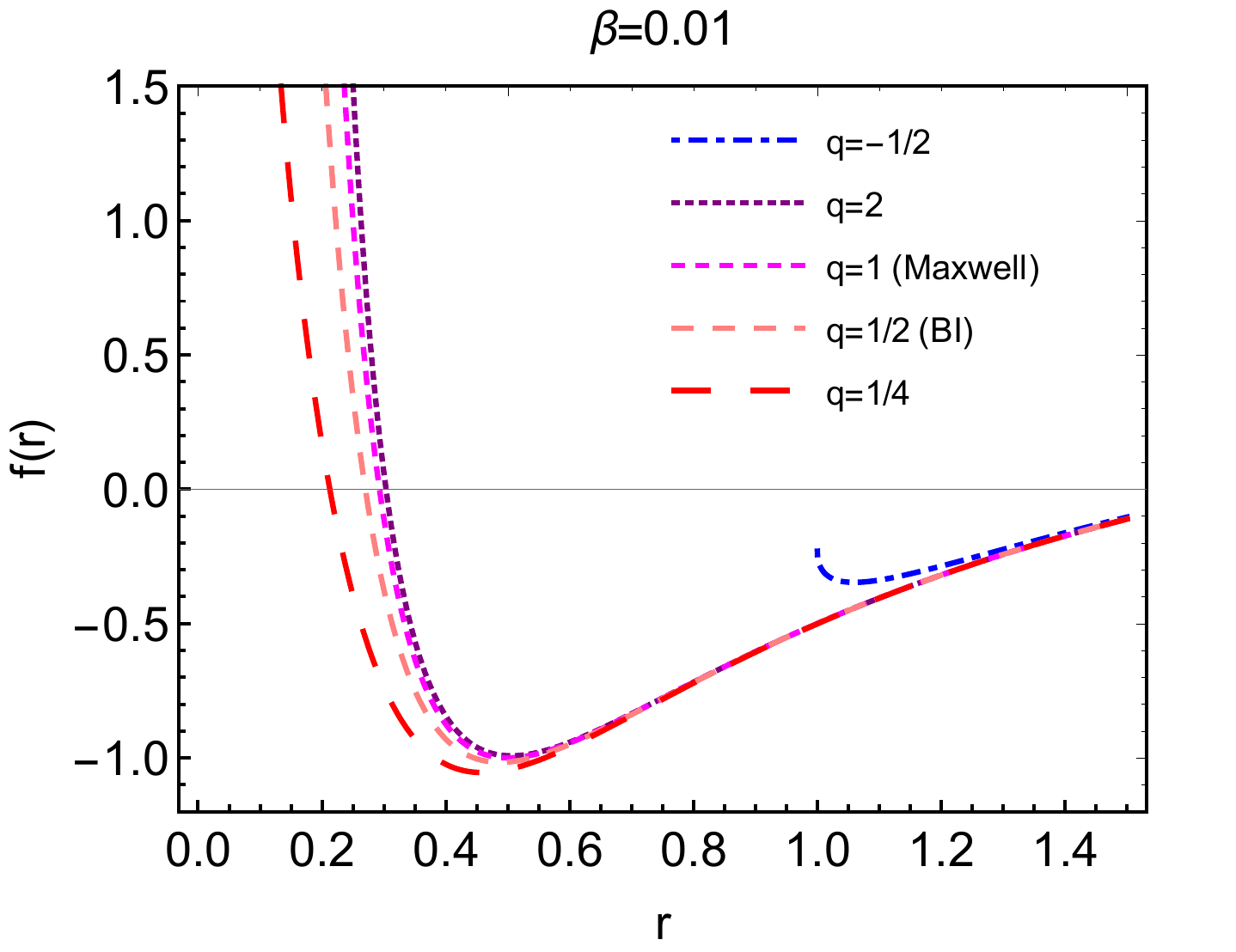} &		\includegraphics[height=6.6cm,keepaspectratio]{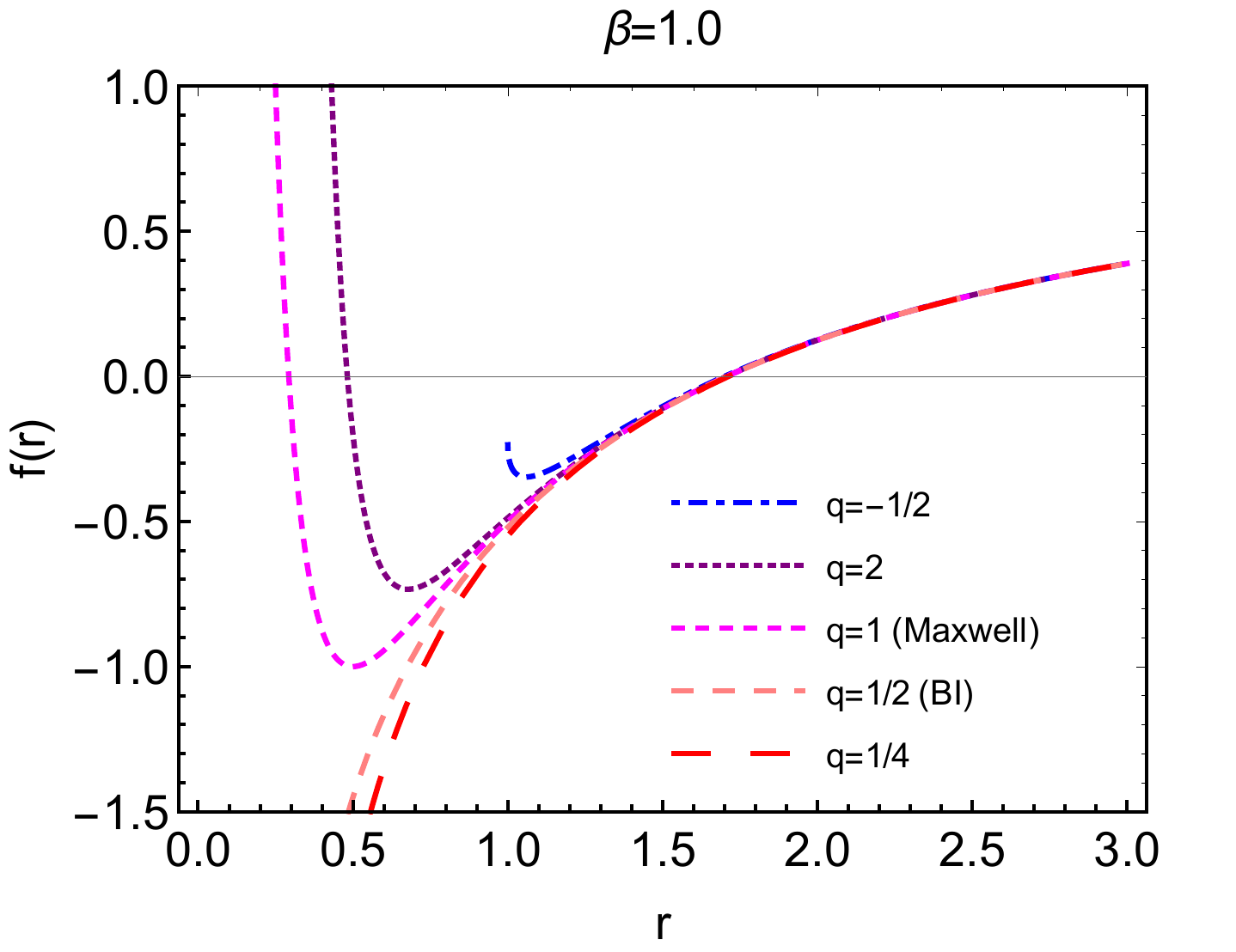}  
	\end{tabular}
	\caption{Typical plots of $f(r)$ with $M=Q=1$.}
	\label{gr_bi_ms_f}
\end{figure}
 \begin{figure}[!ht]
	\centering
	\begin{tabular}{cc}
		\includegraphics[height=6.3cm,keepaspectratio]{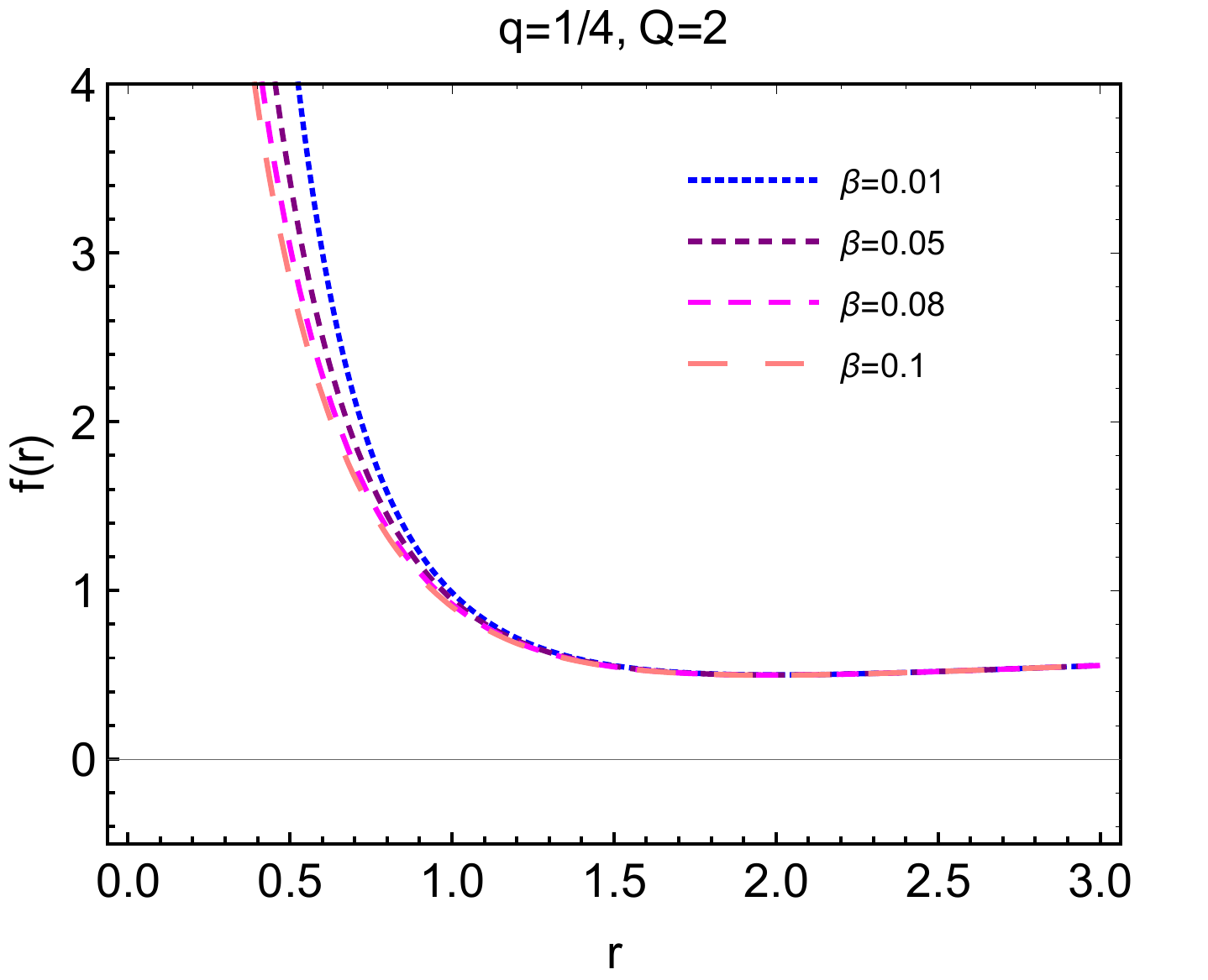} &		\includegraphics[height=6.6cm,keepaspectratio]{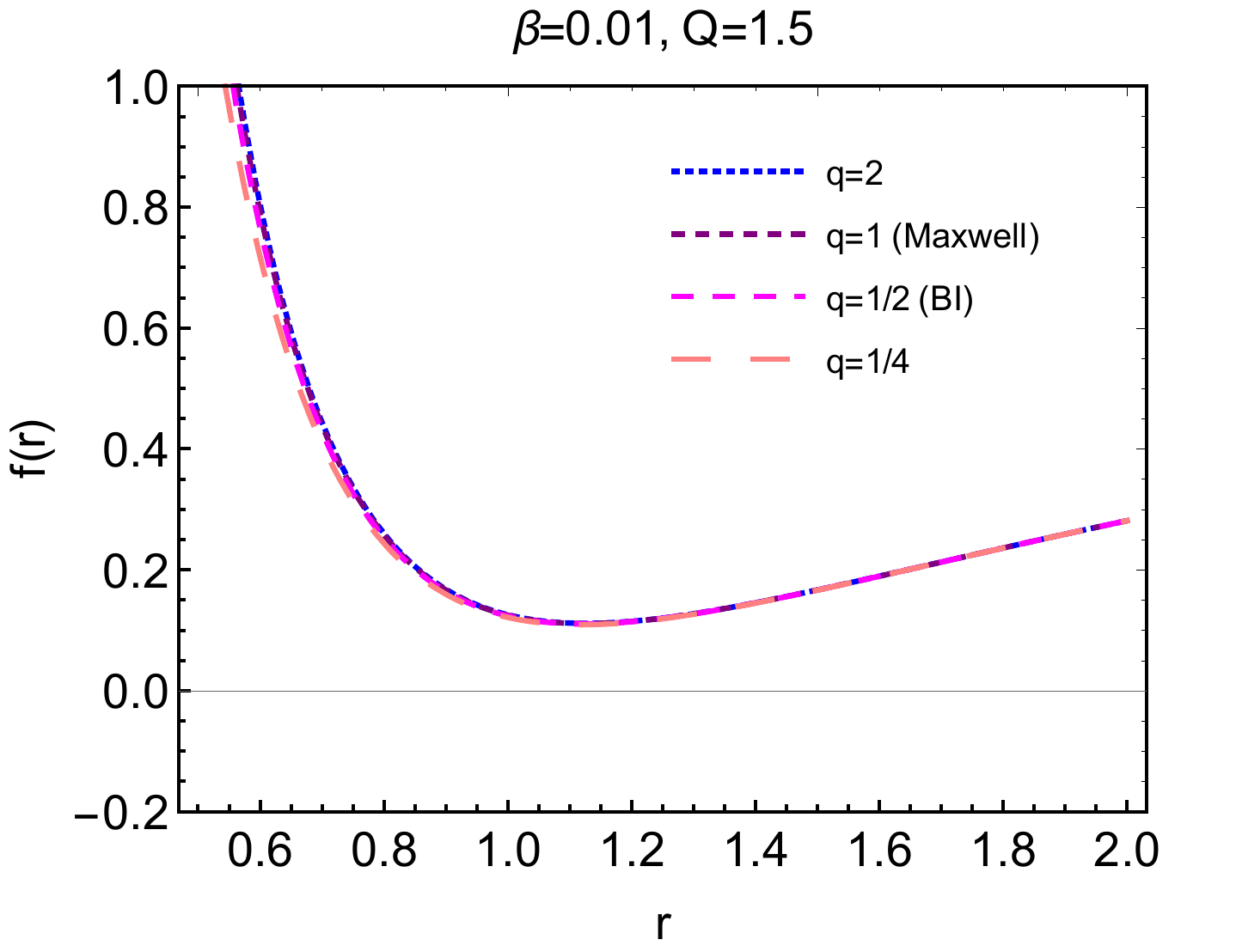}  
	\end{tabular}
	\caption{The cases naked singularities of the metric function $f(r)$. [Left] The no-horizon solution with fixed $q$ for several values of $\beta$. [Right] The vice versa. Here we set $M=1$.}
	\label{gr_bi_ms_f_ns}
\end{figure}
 \begin{figure}[!ht]
	\centering
	\begin{tabular}{cc}
		\includegraphics[height=6.3cm,keepaspectratio]{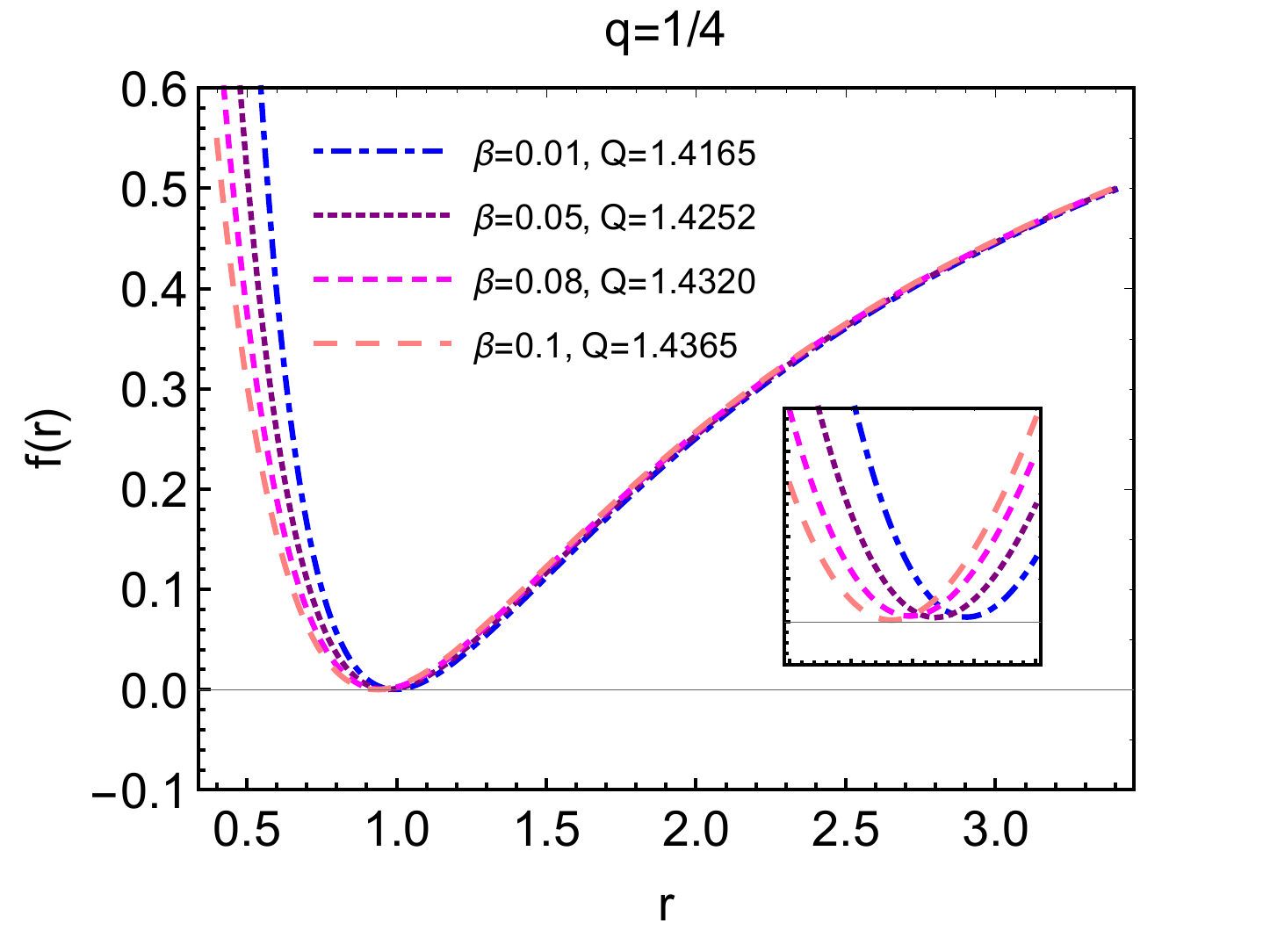} &		\includegraphics[height=6.3cm,keepaspectratio]{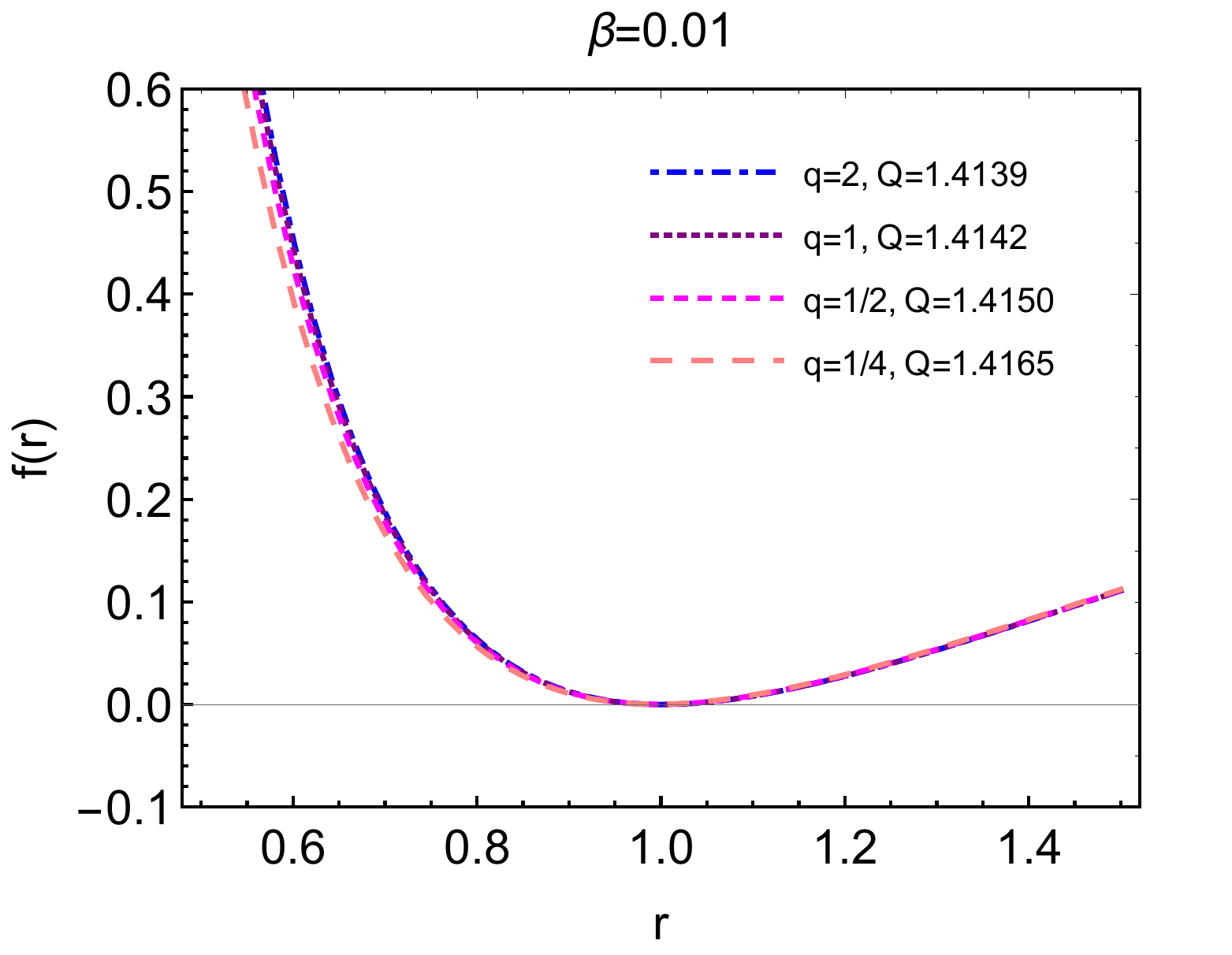}  
	\end{tabular}
	\caption{Typical of extremal cases of the metric function $f(r)$. [Left] The extremal solution with fixed $q$ for several values of $\beta$. [Right] The vice versa. Here we set $M=1$.}
	\label{gr_bi_ms_f_ext}
\end{figure}

In Fig.~\ref{gr_bi_ms_f} we show typical plots of the metric function for several values of $q$, both in the strong and the weak coupling regimes. The behavior does not differ much from the RN solution; they all typically have two horizons. For values of $q$ that does not reduce to Maxwell (for example $q=-1/2$) the metric stops being real. Since they generally possess two horizons, in principle the metric~\ref{gbi_f2r} can be extremal. While it is impossible to show the extremal condition for $M$ and $Q$ analytically, in Fig.~\ref{gr_bi_ms_f_ext} we show that the extremal conditions can be satisfied for certain values of the parameters. The radius tangent to the minima of the metric is the extremal horizon $r_{EH}$. As $\beta$ goes stronger $r_{EH}$ shifts {\it closer} to the singularity. However, for $q<1$ there is a critical value of $\beta$ above which only one horizon exists. This can be seen in Fig.~\ref{gr_bi_ms_f} on the right.

\subsection{Timelike Geodesics}

A test particle with mass $\mu$ and (electric/magnetic) charge $\epsilon$ around compact object can be described by the geodesics equation \cite{Breton:2001yk}
\begin{equation}
\label{geodesic}
\frac{d^2 x^{\nu}}{d\tau^2} +\Gamma_{\alpha\beta}^{\nu} \frac{dx^{\alpha}}{d\tau} \frac{dx^{\beta}}{d\tau} = -\frac{\epsilon}{\mu} F_{\sigma}^{\,\,\nu} \frac{dx^{\sigma}}{d\tau}.
\end{equation}
For our metric \eqref{flat_geo}, the timelike geodesics on equatorial plane ($\theta = \pi/2$) can be written as
\begin{equation}
\label{tl_geo}
1= f \dot{t}^2 -f^{-1} \dot{r}^2 -r^2 \dot{\phi}^2.
\end{equation}
The symmetry of the metric admits conserved quantities
\begin{equation}
\label{dot_E}
\dot{t} = \frac{\mathbb{E}}{f} \,\,\,\, , \,\,\,\,
\dot{\phi} = \frac{\mathbb{L}}{r^2}.
\end{equation}
where $\mathbb{E}$ and $\mathbb{L}$ are the energy-and angular momentum-per unit mass of the test charged particles, respectively. Eq.~\eqref{tl_geo} can be rewritten as
\begin{equation}
\label{timeg_M}
\dot{r}^2 +f \bigg(\frac{\mathbb{L}^2}{r^2} + 1\bigg) -\mathbb{E}^2 = 0.
\end{equation}
Comparing the equation to $\frac{1}{2} \dot{r} + V_{eff} (r) = 0$, we can extract the effective potential as
\begin{equation}
\label{gbi_veff}
V_{eff}(r) = \frac{1}{2} \bigg(\frac{\mathbb{L}^2}{r^2} + 1\bigg) \bigg( 1 -\frac{2 M}{r} -\frac{\kappa ^2 r^2}{3 \beta} \bigg[\, _2F_1\left(-\frac{3}{4},-q;\frac{1}{4};-\frac{Q^2 \beta }{2 q r^4}\right)-1\bigg] \bigg) -\frac{\mathbb{E}^2}{2}.
\end{equation}
It is interesting to note that for monopole black hole massive charged (either electrically or magnetically) test particle behaves the same as the chargeless one.
\begin{figure}
	\centering
	\begin{tabular}{cc}
		\includegraphics[height=6.3cm,keepaspectratio]{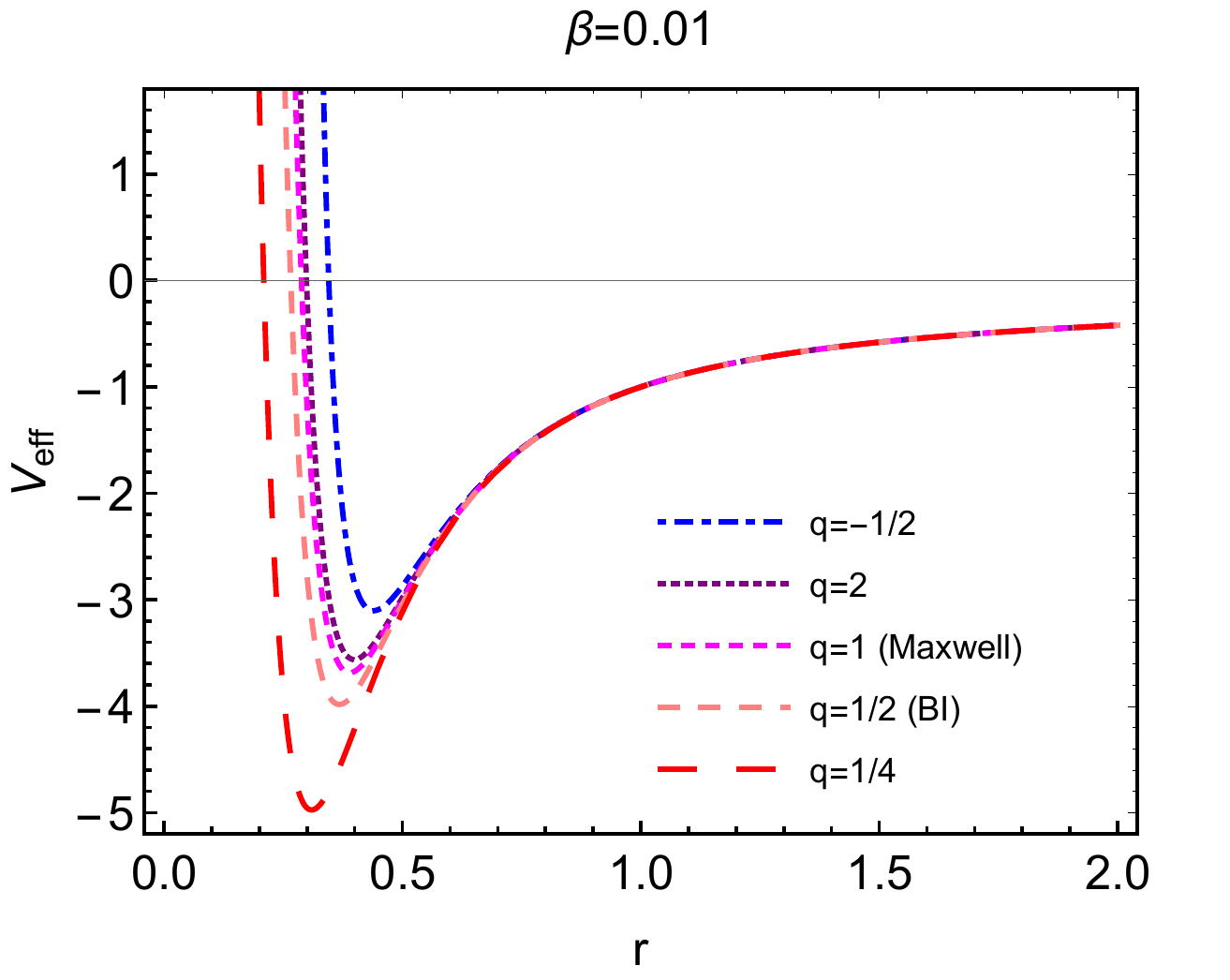} &
		\includegraphics[height=6.3cm,keepaspectratio]{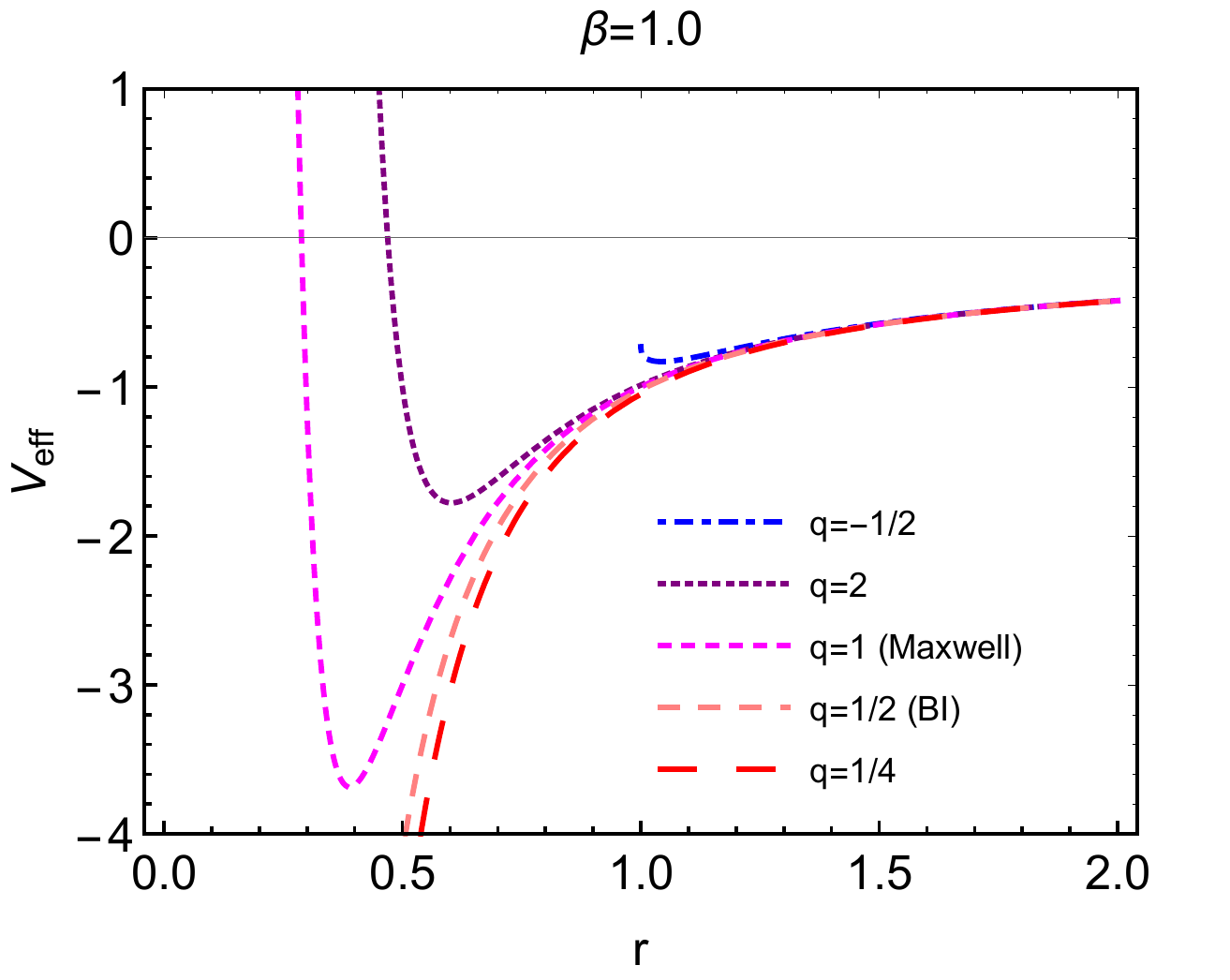}
	\end{tabular}
	\caption{The effective potential for massive particles \eqref{gbi_veff} with $M=Q=1$.}
	\label{gr_bi_ms_geo}
\end{figure}

The plot of $V_{eff}(r)$ for several values of $q$ are shown in Fig.~\ref{gr_bi_ms_geo}. Here we set $\mathbb{E}=\mathbb{L}=1$. The feature of $V_{eff}(r)$ is qualitatively the same as the Newtonian counterpart; there exists bounded orbits. The minimum of $V_{eff}(r)$ corresponds to the radius of {\it stable circular orbit}\footnote{To be precise, here $r_{SCO}=r_{SCO}\left(M, Q, q, \beta\right)$. The smallest $r_{SCO}$ corresponds to the minimum of hypersurface $r_{SCO}$ is the $\left(M, Q, q, \beta\right)$-hyperspace, and is called {\it the Innermost Stable Circular Orbit}, $r_{ISCO}$.} $r_{SCO}$, while its local maximum represents the radius of unstable circular orbit $r_{UCO}$. These closed orbits is constrained by $q$ and the nonlinear coupling $\beta$. For $q\leq-1/2$ the $V_{eff}(r)$ stops being real.


\subsection{Null Geodesics}

Novello {\it et al} showed that in NLED photon follows the null geodesic of its {\it effective} geometry given by~\cite{Novello:1999pg}
\begin{equation}
g^{\mu\nu}_{eff}=\mathcal{L_F}g^{\mu\nu}-4\mathcal{L_{FF}}F^{\mu}_{\ \alpha}F^{\alpha\nu}.
\end{equation}
For our case, it is given by
\begin{equation}
\label{gbi_eff}
g^{\mu\nu}_{eff} = \left( 1+\frac{\beta F}{ q} \right) g^{\mu\nu} - \frac{4\beta (q-1)}{q} F^{\mu\alpha} F_{\alpha}^{\,\,\nu}.
\end{equation}

Defining a factor $h(r)\equiv\frac{2 q r^4+\beta  Q^2}{\beta  (8 q-7) Q^2+2 q r^4}$, the {\it conformally-rescaled} effective line element can be written as
\begin{equation}
\label{gbi_effg}
ds^2_{eff} = - f(r) dt^2+ f(r)^{-1} dr^2+ h(r) r^2 d\Omega^2.
\end{equation}
The null rays in this line element follow the trajectories given by
\begin{equation}
\label{null_deff}
0= f \dot{t}^2 -f^{-1} \dot{r}^2 -hr^2 \dot{\phi}^2,
\end{equation}
from which the $V_{eff}$ can be extracted out as
\begin{equation}
\label{GR_V_effg}
V_{eff} (r)= \frac{f \mathbb{L}^2}{2 h r^2} -\frac{\mathbb{E}^2}{2},
\end{equation}
or, explicitly,  
\begin{equation}
\label{gbi_nullgeo}
V_{eff}(r) = \frac{\mathbb{L}^2}{2 r^2} \bigg( \frac{\beta (8 q-7) Q^2+2 q r^4}{2 q r^4+\beta  Q^2} \bigg) \bigg( 1 -\frac{2 M}{r} -\frac{\kappa ^2 r^2}{3 \beta} \bigg[\, _2F_1\left(-\frac{3}{4},-q;\frac{1}{4};-\frac{Q^2 \beta }{2 q r^4}\right)-1\bigg] \bigg) -\frac{\mathbb{E}^2}{2}, \\
\end{equation}
the stationary of which corresponds to the existence of photon orbits; {\it i.e.,} $V_{eff}'(r)=0$. 
\begin{figure}
	\centering
	\begin{tabular}{cc}
		\includegraphics[height=6.3cm,keepaspectratio]{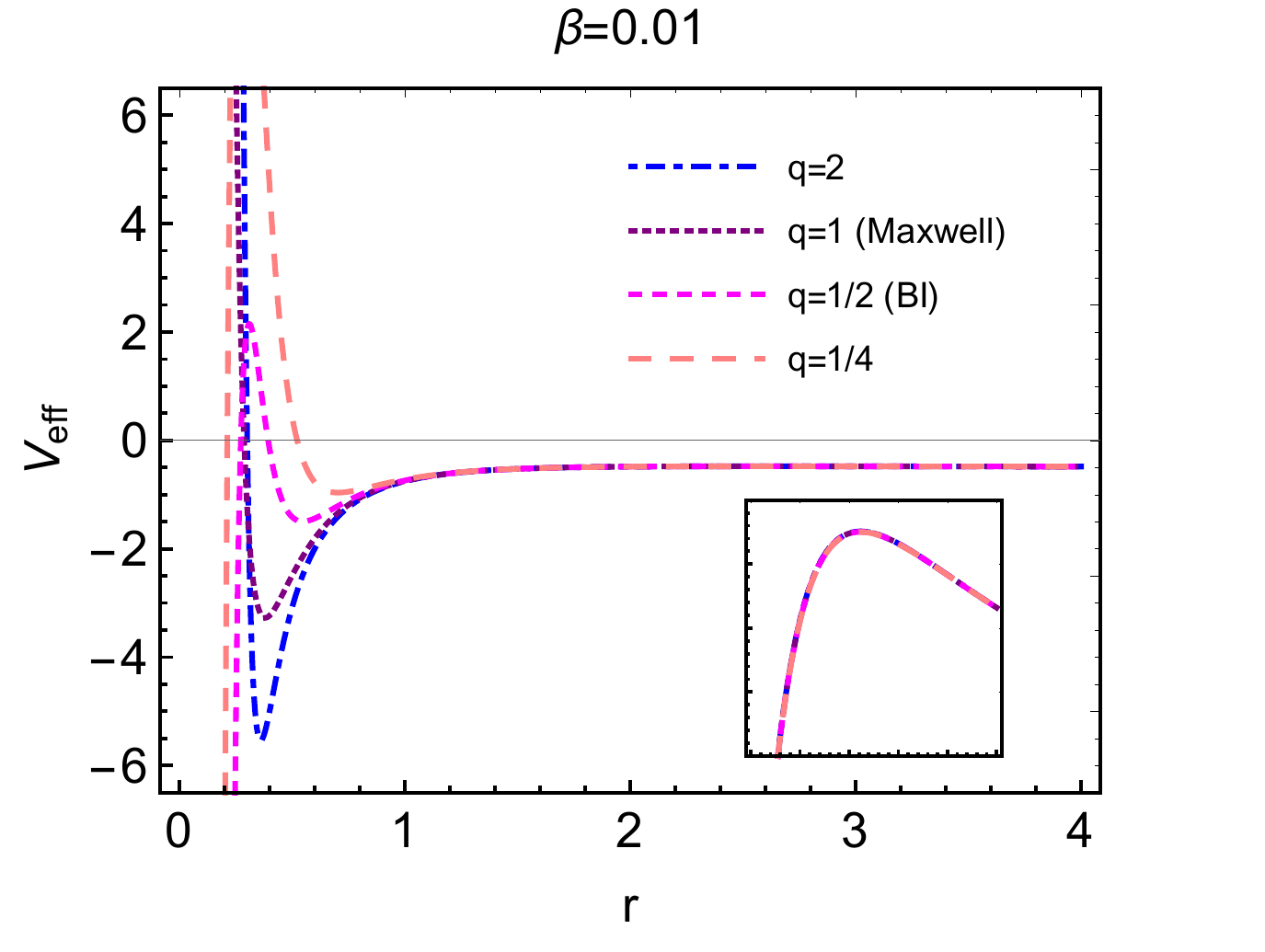} &
		\includegraphics[height=6.3cm,keepaspectratio]{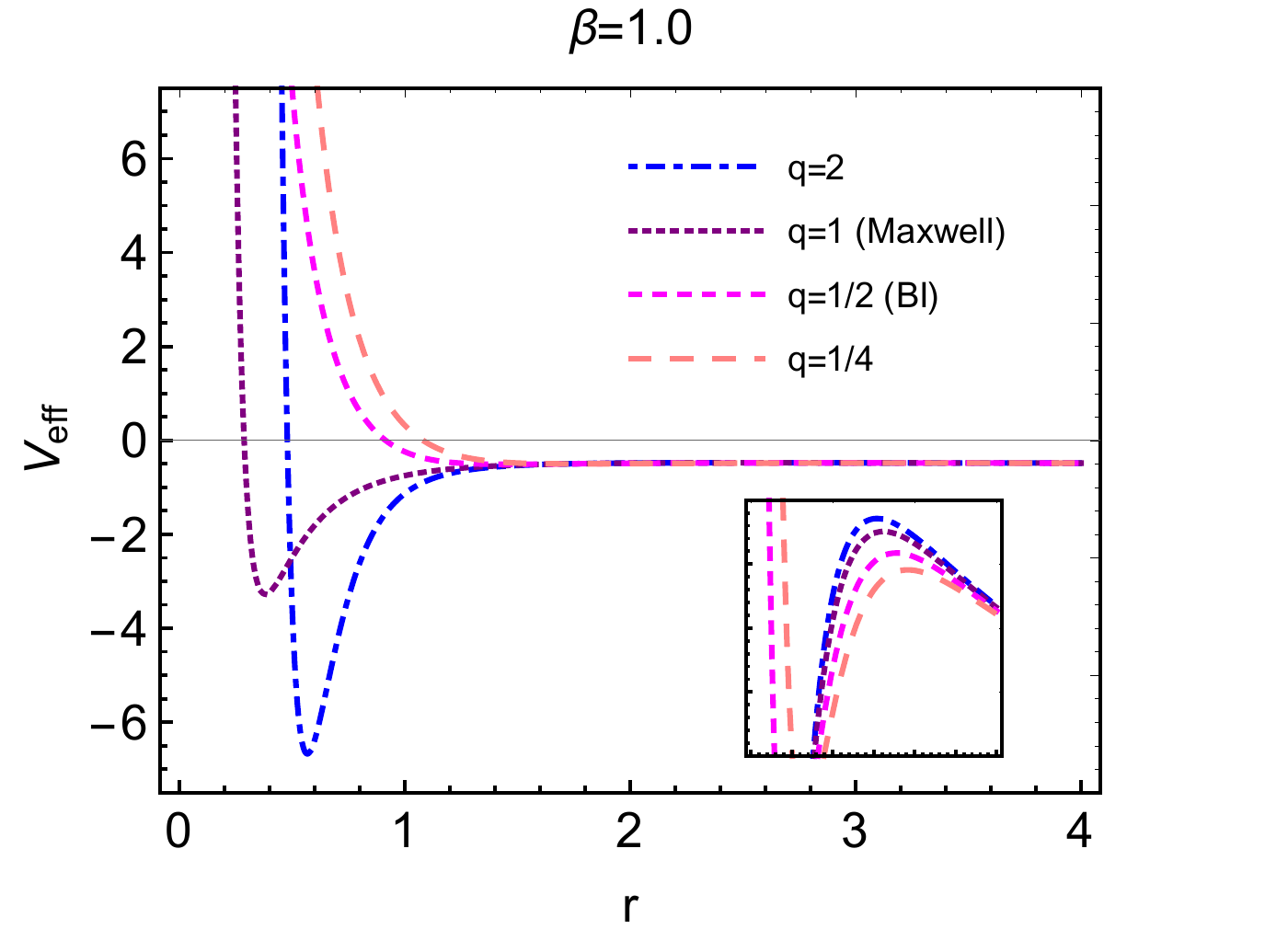}
	\end{tabular}
	\caption{The effective potential for massless particles \eqref{gbi_nullgeo} in the non-extremal case. Here we set $M=Q=1$.}
	\label{gr_bi_ms_null}
\end{figure}

\begin{figure}
	\centering
	\begin{tabular}{cc}
		\includegraphics[height=6.3cm,keepaspectratio]{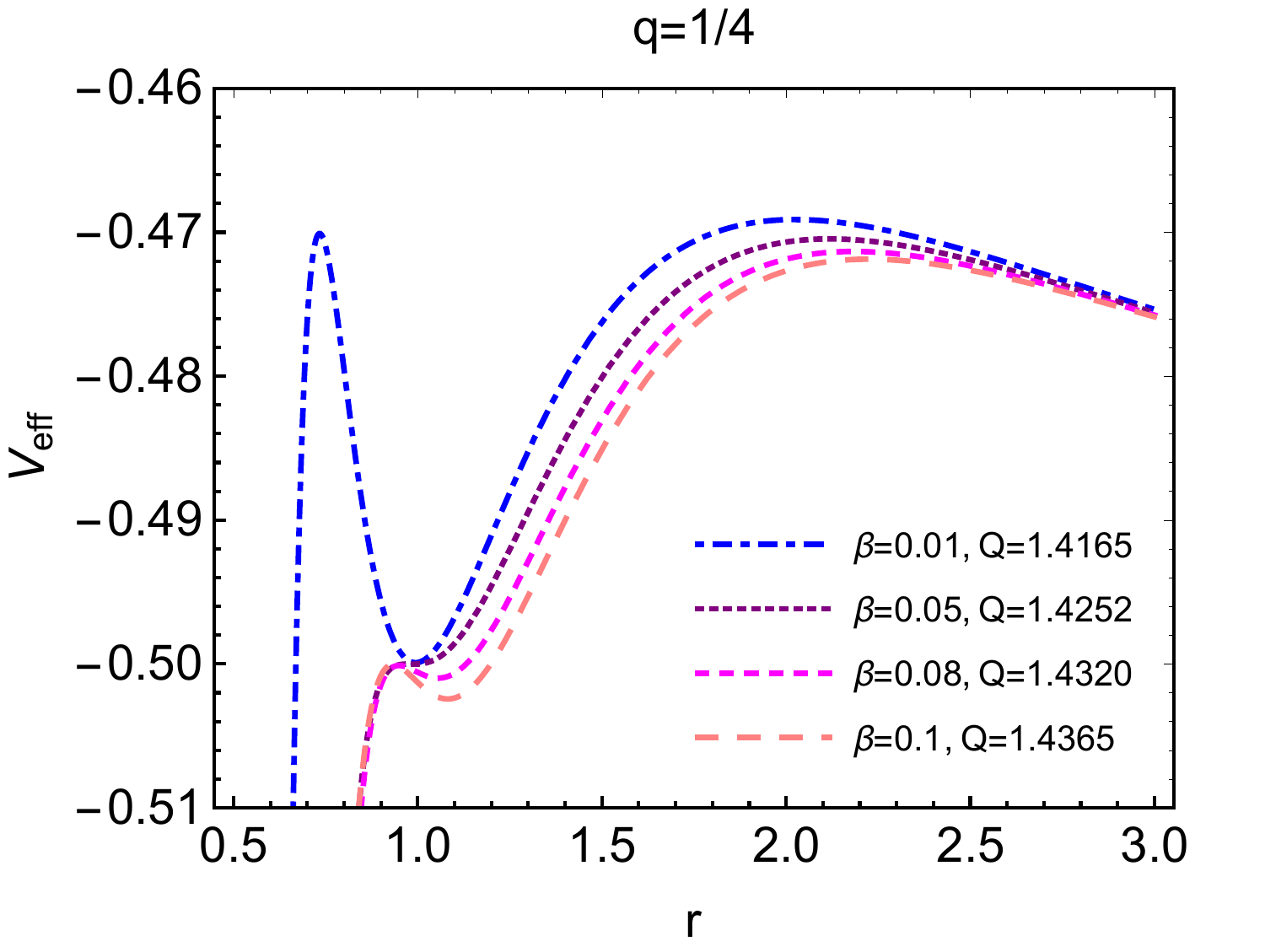} &
		\includegraphics[height=6.3cm,keepaspectratio]{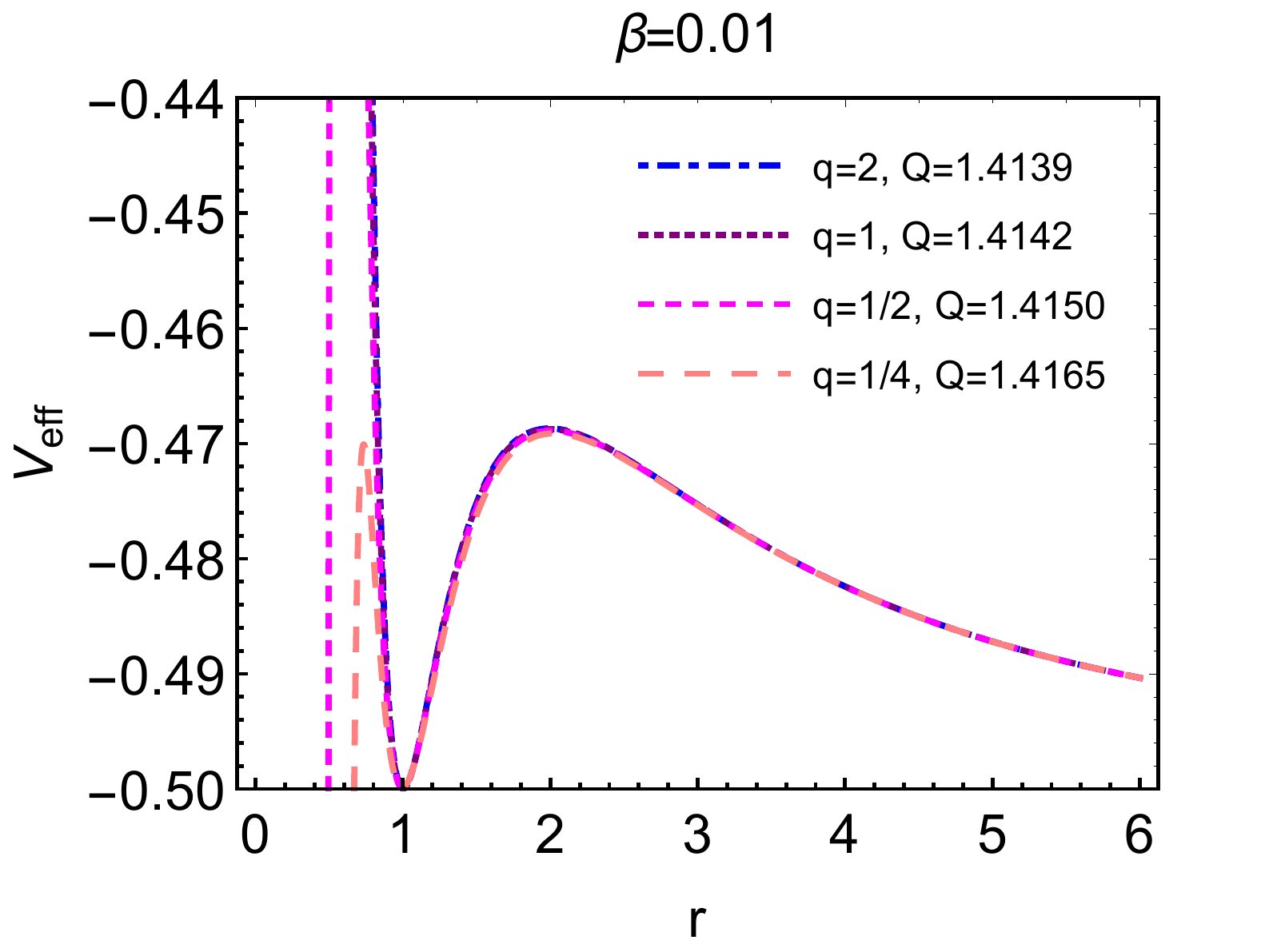}
	\end{tabular}
	\caption{The extremal case of effective potential for massless particles. [Left] $V_{eff}$ with a fixed $q$ and several values of $\beta$ parameter. [Right] The vice versa. We set $M=1$.}
	\label{gr_bi_ms_null_ext}
\end{figure}

\begin{figure}
	\centering
	\includegraphics[height=7cm,keepaspectratio]{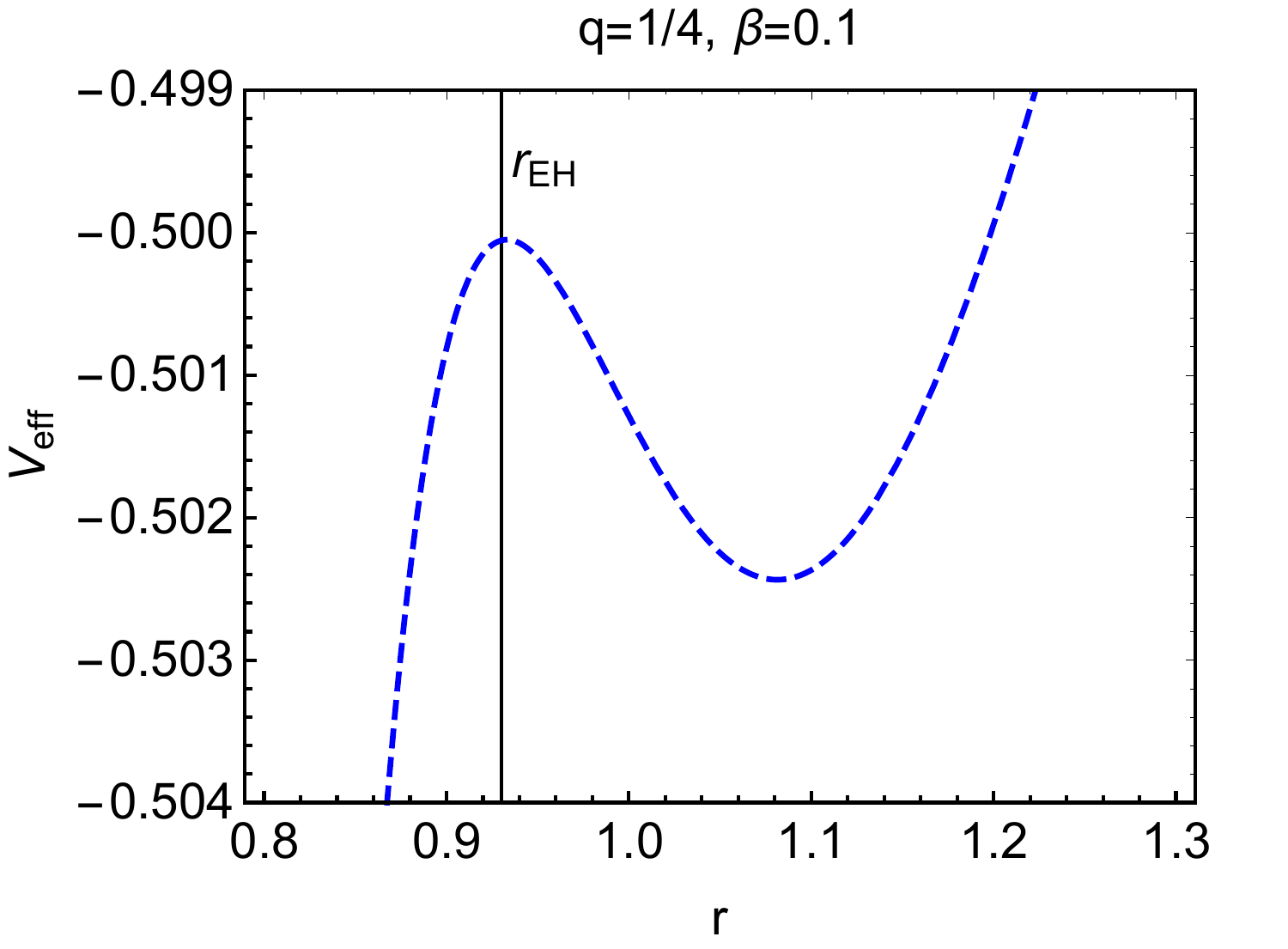}
	\caption{The zoomed-in version of $V_{eff}$ for $q=1/4$ shown in Fig.~\ref{gr_bi_ms_null_ext}. It can be observed that the inner $r_{UCO}$ coincides with $r_{EH}$ (as can be confirmed by Table~\ref{gbi_rEH_beta}) while $r_{SCO}$ lies outside.}
	\label{gr_bi_ms_null_b}
\end{figure}

Since the metric function is not in a simple closed-form function it is of little interest to determine  the $r_{UCO}$ and $r_{SCO}$ analytically. In Figs.~\ref{gr_bi_ms_null}-\ref{gr_bi_ms_null_ext} we plot $V_{eff}(r)$ for photon. In the non-extremal case the situation is similar as in RN, except in the weak-coupling regime and $q<1$ we have $r_{UCO}<r_{SCO}$. Both are still inside the corresponding outer horizon. For the extremal case, however, something interesting emerges. As can be seen from Fig.~\ref{gr_bi_ms_null_ext} there are typically two $r_{UCO}$ and one $r_{SCO}$, and the position of $r_{SCO}$ shifts {\it farther away} from singularity when $\beta$ gets stronger, as opposed to the behavior of $r_{EH}$ discussed earlier. As a result, there is a window of parameter space where you can have stable photon orbits {\it outside} the extremal horizon; {\it i.e.,} $r_{SCO}>r_{EH}$. This is clearly shown in Fig.~\ref{gr_bi_ms_null_b}. The consequence is novel. Not only stable circular photon orbit is possible, but there exists a family of bounded orbits with $r_{min}\leq r\leq r_{max}$, as long as $r_{min}\geq r_{EH}$. 

\begin{figure}
	\centering
	\begin{tabular}{ccc}
		\includegraphics[height=6cm,keepaspectratio]{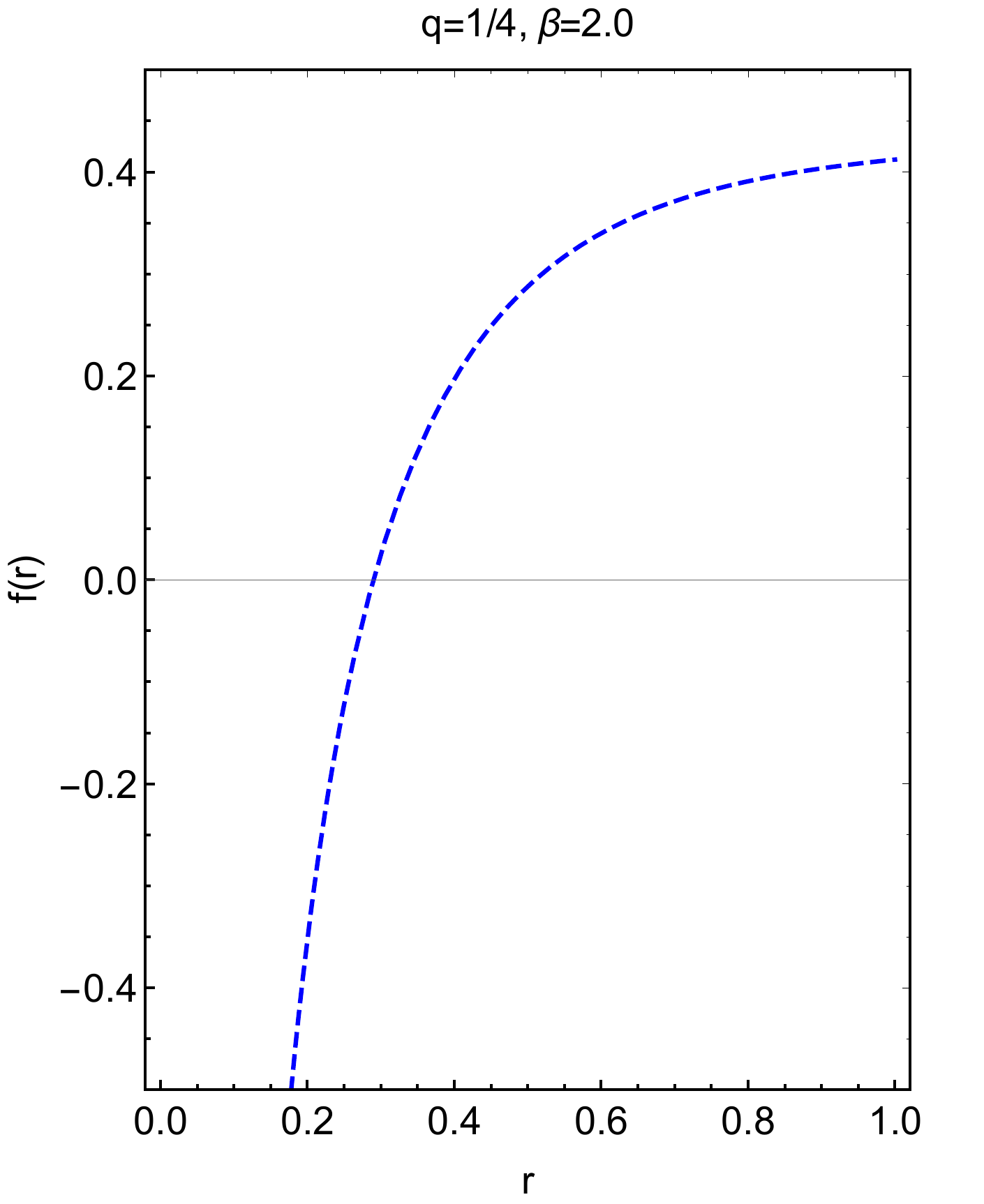} &
		\includegraphics[height=6cm,keepaspectratio]{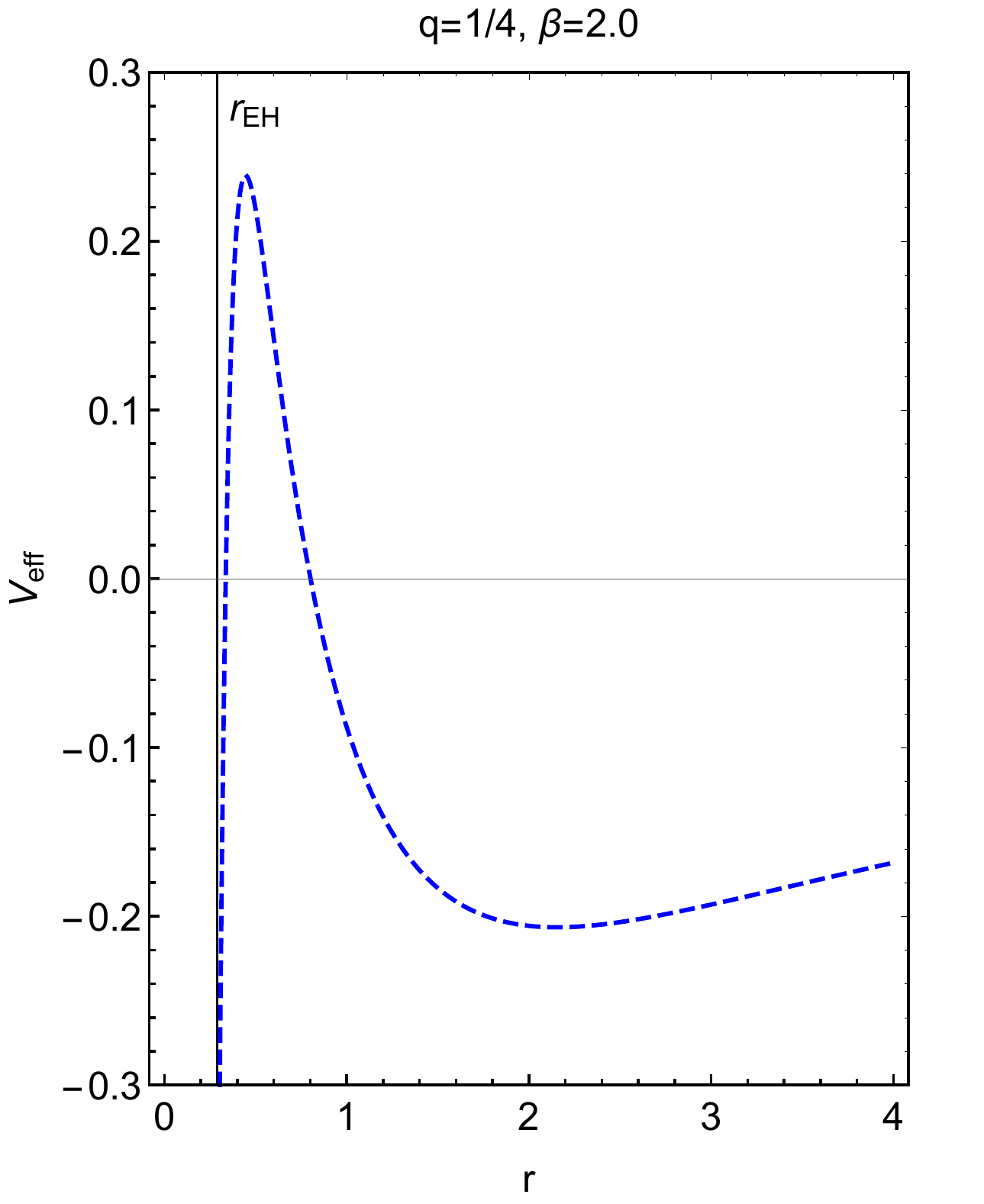} &
		\includegraphics[height=6cm,keepaspectratio]{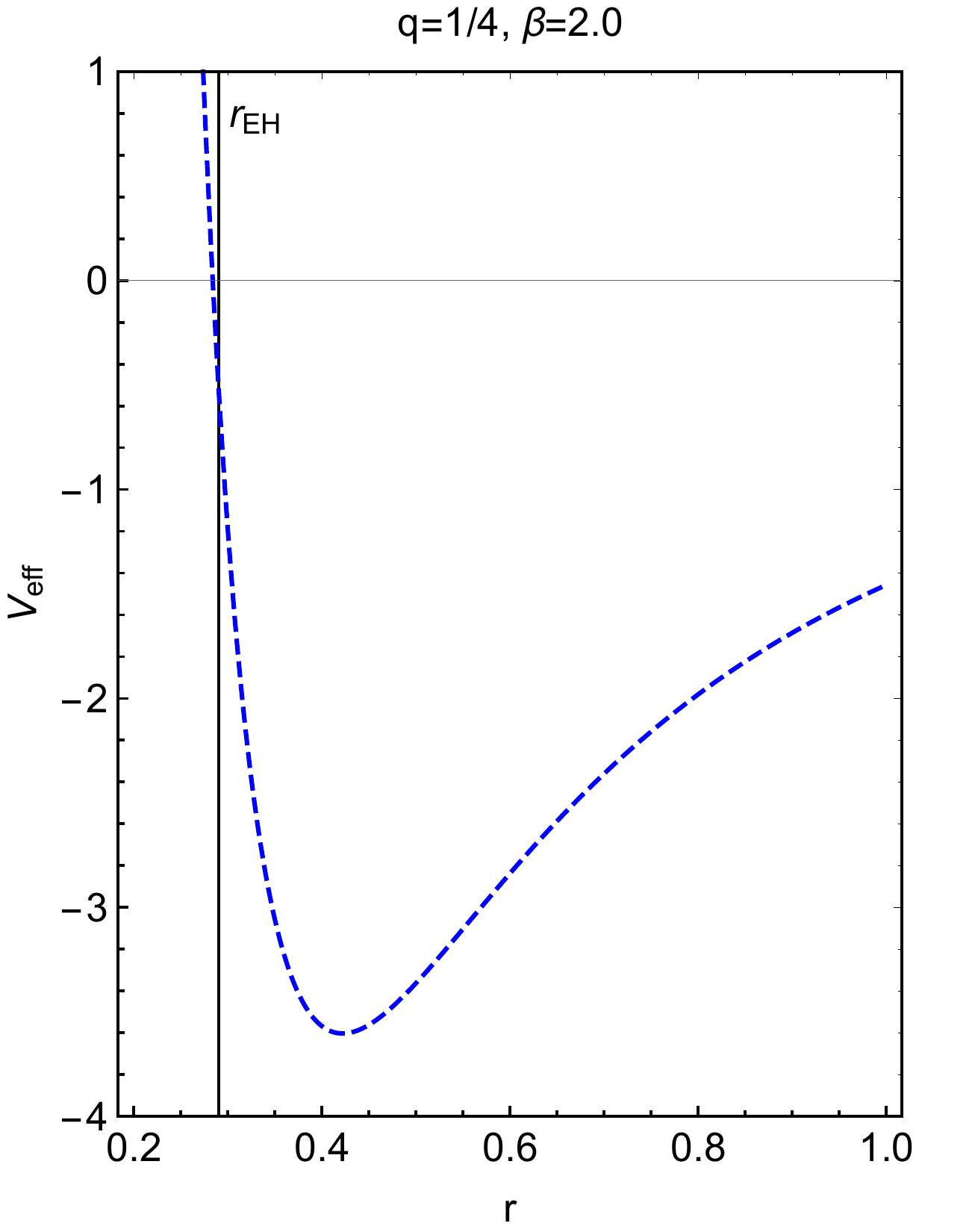}
	\end{tabular}
	\caption{The metric function $f(r)$ (left), effective potential for massive particles (center) and effective potential for light particles (right) with $M=1$ and $Q=2$.}
	\label{gr_bi_geo_b2}
\end{figure}

As we mentioned in the previous subsection, for $q<1$ the inner horizon ceases to exist as $\beta$ gets stronger; {\it i.e.,} the black hole behaves Schwarzschild-like. The relevant thing is whether $r_{SCO}$ is located inside the remaining single horizon $r_h$. In Fig.~\ref{gr_bi_geo_b2} we plot the metric (left panel) as well as its corresponding $V_{eff}$ for timelike (center panel) and null (right panel) in the strong-coupling limit ($\beta=2$) and $q=1/4$. It can be observed that timelike $V_{eff}$ possesses $r_{UCO}$ and $r_{SCO}$ with $r_{UCO}<r_{SCO}$. On the other hand, for null geodesic only $r_{SCO}$ exists, and $r_{h}<r_{SCO}$. Thus for $\beta=2$, and we infer that it is valid for strong-coupling regime $\beta\geq1$, stable photon orbits also exist. In Table.~\ref{gbi_rEH_beta} we show physical photon orbits with $r_{SCO}>r_{EH}$ for the case of $q=1/4$ with varying $\beta$ and the values of BH charge $Q$. This is a typical family of solutions with $q<1$. For the specific $\beta=0.05$  we found that the $r_{SCO}$ is rather metastable since it is the saddle point of $V_{eff}$; {\it i.e.,} $V''(r_{SCO})=0$.
 
\begin{table}
\setlength{\tabcolsep}{1em}
	\begin{tabular}{||c | c | c | c | c||} 
	\hline
	$\beta$ & $Q$ & $r_{EH}$ & $r_{UCO}$ & $r_{SCO}$ \\ [0.5ex] 
	\hline\hline
	2.0 & 2.0 & 0.2900 & - & 0.4220 \\ 
	\hline
	0.1 & 1.4365 & 0.9300 & 0.934 and 2.223 & 1.081 \\ 
	\hline
	0.08 & 1.4320 & 0.9475 & 0.951 and 2.187 & 1.054 \\
	\hline
	0.05 & 1.4252 & 0.9690 & 2.117 & 0.980* \\
	\hline
	0.01 & 1.4165 & 0.9899 & 0.734 and 2.023 & 0.994 \\
	\hline
	\end{tabular}
	\caption{Comparation of the radius of event horizon of the extremal case ($r_{EH}$) with $q=1/4$ for various number of $\beta$, with its $r_{SCO}$ and $r_{SCO}$. The starred value is a saddle point.}
	\label{gbi_rEH_beta}
\end{table}

\subsection{Null Geodesics in Born-infeld case ($q=1/2$)}

\begin{figure}
	\centering
	\begin{tabular}{cc}
		\includegraphics[height=6.3cm,keepaspectratio]{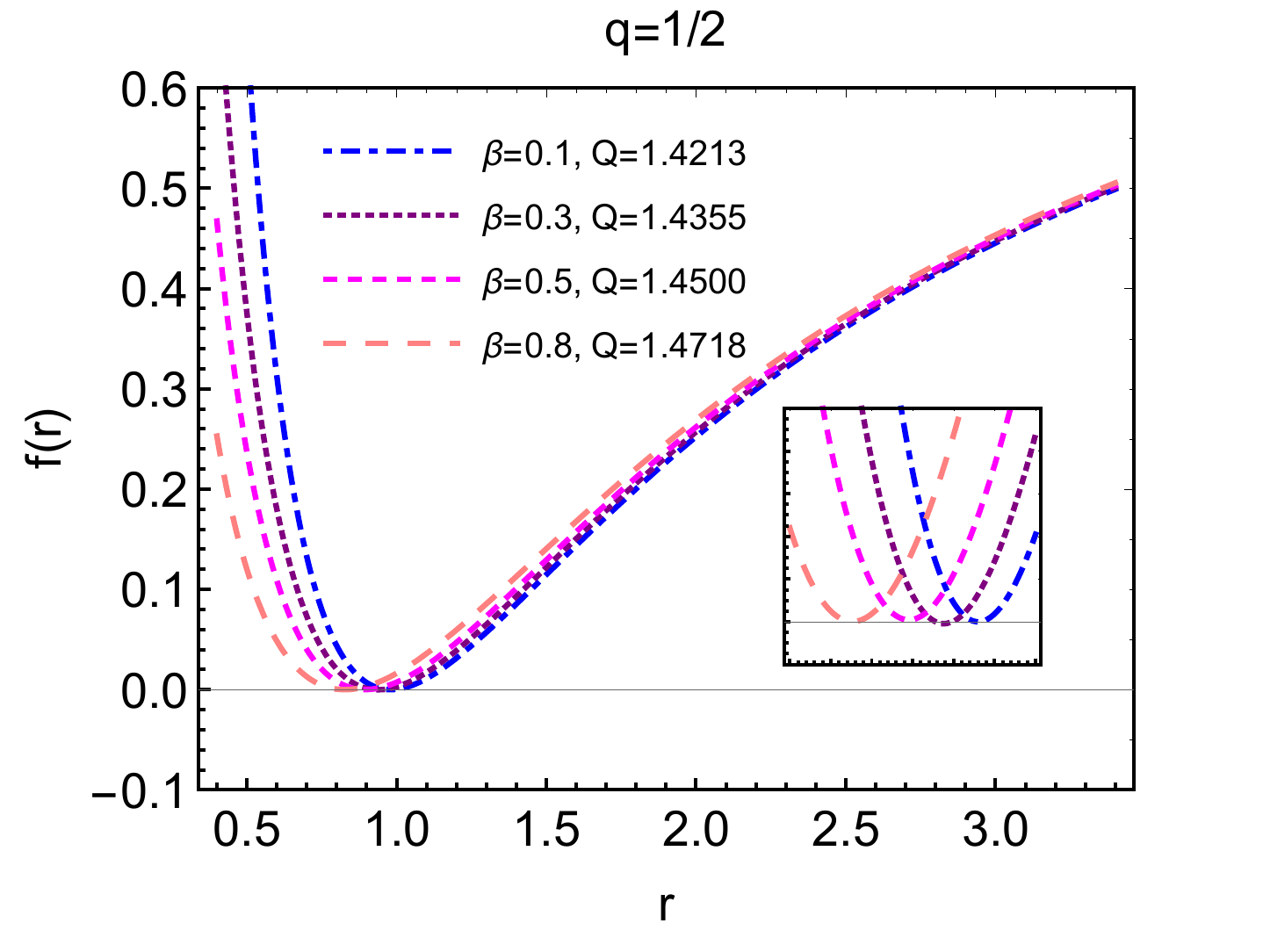} &
		\includegraphics[height=6.3cm,keepaspectratio]{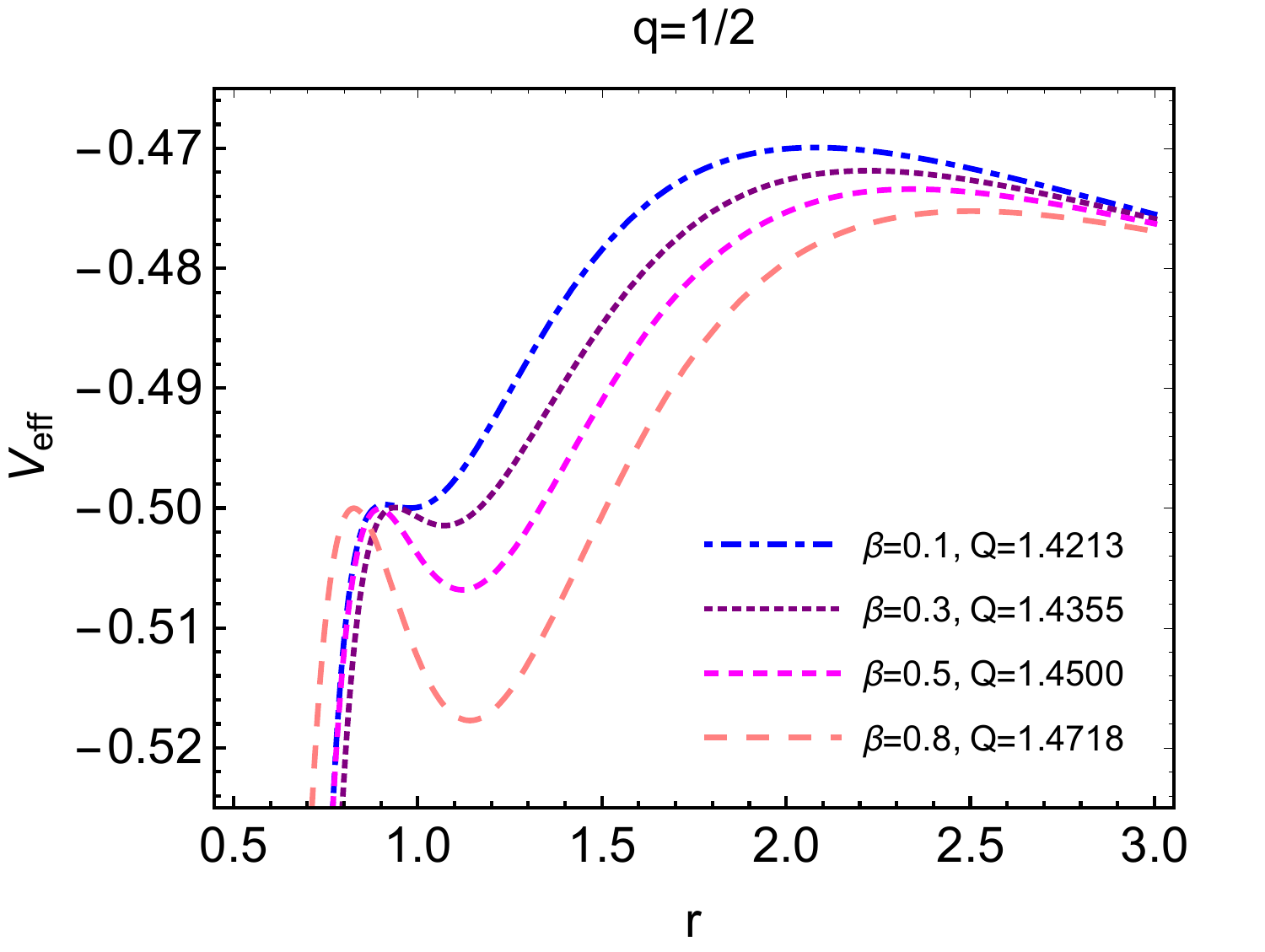}
	\end{tabular}
	\caption{[Left] The extremal case of metric function $f(r)$ and; [Right] The extremal case of effective potential for massless particles with $q=1/2$ and several values of $\beta$ parameter. Here we set $M=1$.}
	\label{bi_ms_null_ext}
\end{figure}

In this section we examine the null behaviour of $q=1/2$ which reduces our model to BI electrodynamics. While the null geodesics of BI model has been studied in the past \cite{Breton:2001yk,Fernando:2012cg}, there is only a few extensive studies that is worked on the magnetostatic scenario \cite{Breton:2005ye, Breton:2007bza}. It is worth to note that the BI enjoys $SO(2)$-duality invariance (not necessarily present in other NLED), where the spherically symmetric solution is \textit{exactly} the same for electric and magnetic case. Breton has shown in her paper that for electric BI blackhole possesses $r_{SCO}$ outside the $r_{EH}$ \cite{Breton:2001yk}. Through our study, we want to take a look if the magnetic case also performs the same result. Here we evaluate the solution in extremal case. The metric function $f(r)$ and its potential $V_{eff}$ is shown in Fig. \ref{bi_ms_null_ext}. 

Here we see the metric behaviour is similar to the previous case ($q=1/4$). The event horizon radius $r_{EH}$ gets smaller as the value of $\beta$ increases. The potential, on the other hand, shows that the $r_{SCO}$ moves \textit{farther away} from the center of the black holes as $\beta$ rises. We analyze the numbers and we find that the $r_{SCO}$ lies outside the event horizon for almost all value of $\beta$, with the case of $\beta=0.1$ has the $r_{EH}$ and $r_{SCO}$ coincide. In Table.~\ref{bi_rEH_null} we show physical photon orbits with $r_{SCO}>r_{EH}$ for the case of $q=1/2$ with varying $\beta$ and the values of BH charge $Q$. 

\begin{figure}
	\centering
	\includegraphics[height=7cm,keepaspectratio]{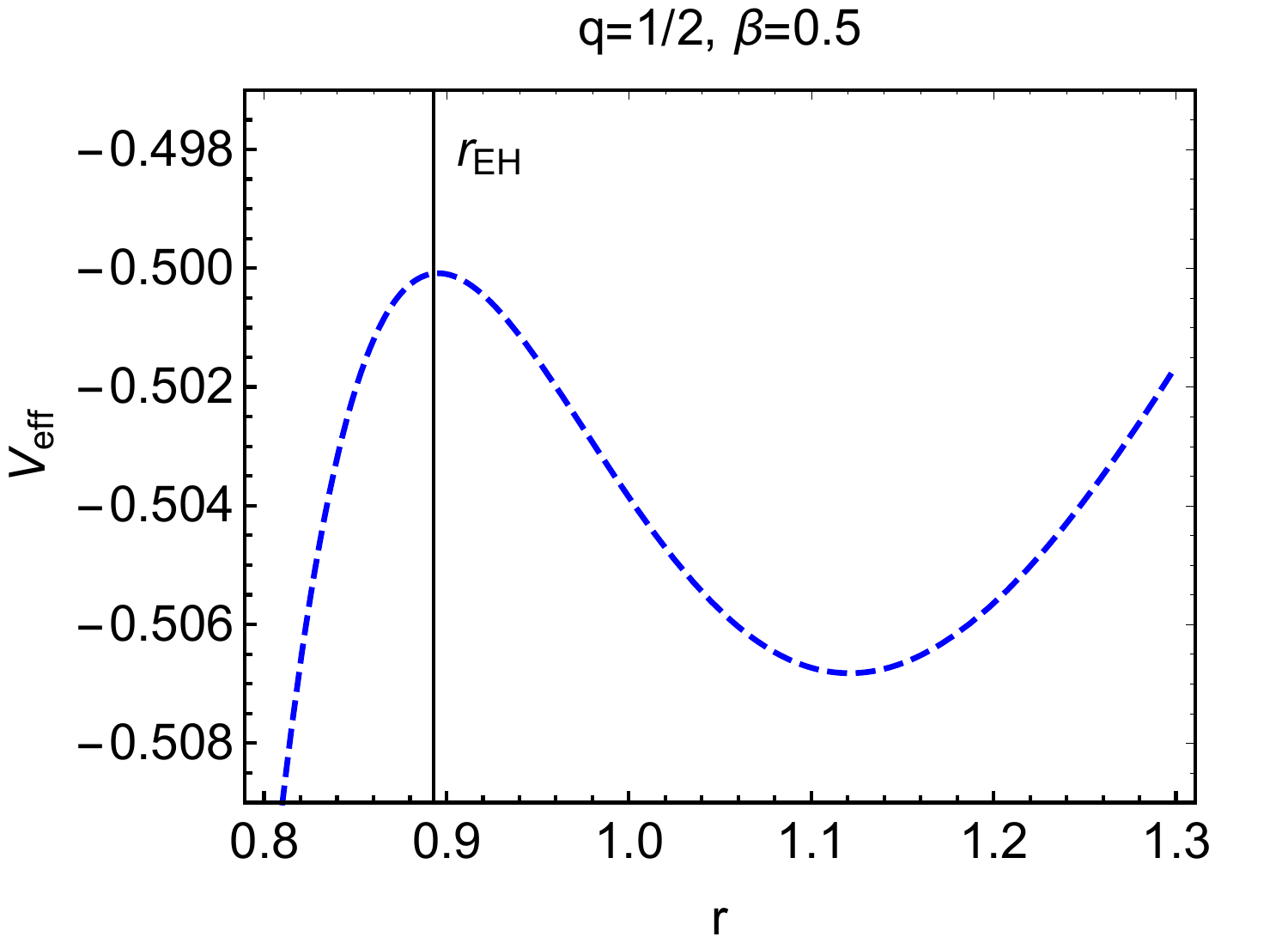}
	\caption{The zoomed-in version of $V_{eff}$ for Born-Infeld case shown in Fig.~\ref{bi_ms_null_ext}. It can be observed that the inner $r_{UCO}$ coincides with $r_{EH}$ (as can be confirmed by Table~\ref{bi_rEH_null}) while $r_{SCO}$ lies outside.}
	\label{bi_ms_null_b}
\end{figure}

\begin{table}
	\setlength{\tabcolsep}{1em}
	\begin{tabular}{||c | c | c | c | c||} 
	\hline
	$\beta$ & $Q$ & $r_{EH}$ & $r_{UCO}$ & $r_{SCO}$ \\ [0.5ex] 
	\hline\hline
	0.1 & 1.4213 & 0.970 & 0.910 and 2.077 & 0.981 \\ 
	\hline
	0.3 & 1.4355 & 0.935 & 0.937 and 2.221 & 1.072 \\
	\hline
	0.5 & 1.4500 & 0.893 & 0.895 and 2.344 & 1.122 \\
	\hline
	0.8 & 1.4718 & 0.830 & 0.827 and 2.502 & 1.142 \\
	\hline
\end{tabular}
\caption{Comparation of the radius of event horizon of the extremal case ($r_{EH}$) in Born-Infeld case for various number of $\beta$, with its $r_{SCO}$.}
	\label{bi_rEH_null}
\end{table}



\subsection{Deflection of Light}

As the last analysis for this model, let us calculate the (weak) deflection angle of light in the case other than BI. Consider the case of $q=1/4$. Knowing the conserved quantities, $\mathbb{E}=f \dot{t}$, $\mathbb{L}=h r^2 \dot{\phi}$ and defining impact parameter $b_0=\mathbb{L}/\mathbb{E}$, we rewrite the null geodesics in term of $u\equiv1/r$ as
\begin{equation}
\label{gbi_null}
\frac{d^2 u}{d \phi ^2} +f h u= -\frac{u^2}{2} \frac{d}{d u} (f h) + \frac{1}{2 b_0^2} \frac{d}{d u} (h^2).
\end{equation}
Assuming small $\beta << 1$, we might expand the metric function $f$ and conformal factor $h$ using Taylor series for first order of $\beta$. Inserting the corresponding function, Eq. \eqref{gbi_null} up to first order in $\beta$ is
\begin{equation}
\label{gbi_null_b2}
\frac{d^2 u}{d \phi ^2} + u= 3 M u -\kappa ^2 Q^2 u^3 +\beta  \left(\frac{48 Q^2 u^3}{b_0^2}+84 M Q^2 u^6-\frac{1}{10} 237 \kappa ^2 Q^4 u^7-36 Q^2 u^5\right).
\end{equation}

Define $\epsilon\equiv M u_0$ and $\xi\equiv u/u_0$. This yields Eq.~\eqref{gbi_null_b2}, up to second-order in $\epsilon$,
\begin{equation}
\label{gbi_null_xi}
\frac{d^2 \xi}{d \phi ^2} + \xi \approx 3 \xi ^2 \epsilon +\xi^3 \epsilon^2 \left(\frac{48 \beta Q^2}{\text{b0}^2 M^2}-\frac{\kappa ^2  Q^2}{M^2}\right).
\end{equation}
Now, expand $\xi$ in power of $\epsilon$, $\xi =\xi_0  +\epsilon \xi_1  + \epsilon^2 \xi_2  + ....$, then insert them into Eq.~\ref{gbi_null_xi}. We can sort the equation by collecting terms in different order of $\epsilon$~\cite{Jana:2015cha}:
\begin{eqnarray}
\label{chi_0}
\frac{d^2 \xi _0}{d \phi ^2}+\xi _0 &=& 0,\nonumber\\
\label{chi_1}
\frac{d^2 \xi _1}{d \phi ^2}-3 \xi _0^2+\xi _1 &=& 0,\nonumber
\\
\label{chi_2}
\frac{d^2 \xi _2}{d \phi ^2} +\frac{\kappa ^2 \xi_0^3 Q^2}{M^2}-\frac{48 \beta  \xi_0^3 Q^2}{b_0^2 M^2} -6 \xi _1 \xi _0+\xi _2 &=&0.
\end{eqnarray}
Solving them, we find the approximation of inverse radial distance $u$ as
\begin{eqnarray}
\label{gbi_u}
u&\simeq& u_0 \cos (\phi ) +\frac{1}{2} M u_0^2 (3-\cos (2 \phi ))\nonumber\\
&&+\frac{u_0^3}{32 b_0^2} \bigg[12 \phi  \sin (\phi ) \left(\text{b0}^2 \left(10 M^2-\kappa ^2 Q^2\right)+48 \beta  Q^2\right)\nonumber\\
&&+\cos (\phi ) \left(\text{b0}^2 \left(54 M^2-7 \kappa ^2 Q^2\right)+336 \beta  Q^2\right)+\cos (3 \phi ) \left(\text{b0}^2 \left(6 M^2+\kappa ^2 Q^2\right)-48 \beta  Q^2\right) \bigg].\nonumber\\
\end{eqnarray}

\begin{figure}[!ht]
	\centering
	\includegraphics[height=8cm,keepaspectratio]{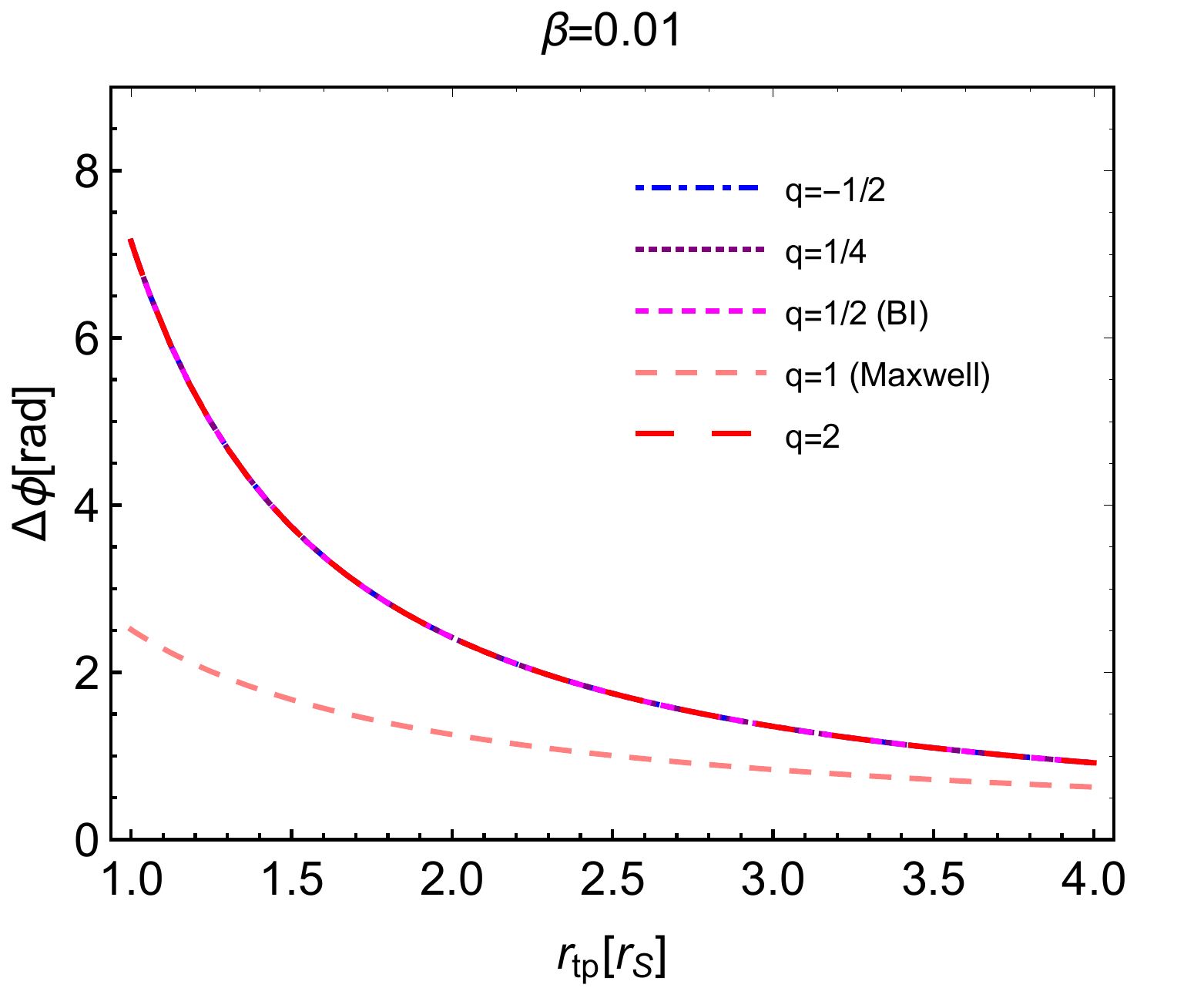} 
	\caption{Deflection angles for various number of $q$.}
	\label{gr_gbi_ms_def}
\end{figure}

Asymptotically $\phi\rightarrow\pi/2 + \delta$ as $u\rightarrow 0$. For the case of $u_0 \approx 0$, we can solve $\delta$ keeping only second order of $u_0$ as
\begin{equation}
\label{gbi_delta}
\delta \approx 2 M u_0 +u_0^2 \left(\frac{9 \pi  \beta  Q^2}{b_0^2}+\frac{15 \pi  M^2}{8}-\frac{3}{16} \pi  \kappa ^2 Q^2\right).
\end{equation}
We want to see the contribution of first order $\beta$ in the deflection angle. We use $b_0\approx1/u_0\equiv r_{tp}$ where $r_{tp}$ is the radius of turning point. The weak deflection angle can be obtained as
\begin{equation}
\Delta \phi_{weak}\equiv2 \delta \approx \frac{4 M}{r_{\text{tp}}} + \frac{15 \pi  M^2}{4 r_{\text{tp}}^2} +Q^2 \left(\frac{18 \pi  \beta }{r_{\text{tp}^4}}-\frac{3 \pi  \kappa ^2}{8 r_{\text{tp}}^2}\right).
\end{equation}
Using the same method, the deflection angle for other cases of $q$ is calculated. The behaviour is shown in Fig.~\eqref{gr_gbi_ms_def}.

In Fig.~\eqref{gr_gbi_ms_def} above we show the behaviour of deflection angles by setting the value of its parameter with $M=10 \,\, M_{\astrosun}$ (solar mass), the charge is set (arbitrarily) to be $Q=0.3$,  and the radius of turning point $r_{tp}=b$ in solar radius $r_{\astrosun}$ which has been normalized by Schwarzchild radius $r_S$. It can be seen that that the other cases beside Maxwell ($q \neq 1$) lay on the same curves. They all asymptote to Maxwell for large $r_{tp}$, as they should, but significantly differ in the short-length regime. The extra term of $\beta$ contributes bigger value of deflection angle near the Schwarzchild radius. Obviously we cannot trust this weak approximation all the way to $r_{tp}=r_S$ since it is the regime where the field gets strong and thus full strong deflection analysis is required~\cite{Bozza:2001xd, Bozza:2002zj, Eiroa:2010wm, Eiroa:2005ag}.

\section{Power-law NLED}
\label{GM_GR}

This power-law electrodynamics was proposed by Hassaine and Martinez~\cite{Hassaine:2008pw, Hassaine:2007py} and is given by
\begin{equation}
\label{gm_nled}
\mathcal{L} = -\mathcal{F}^q.
\end{equation}
Maxwell is recovered when $q=1$. The corresponding ``permittivity" and ``permeability" are expressed as
\begin{equation}
\label{gm_eps_mu}
\epsilon_m=\mu_m^{-1}=q \mathcal{F}^{q-1}.
\end{equation}
They found that the BH is given by 
\begin{eqnarray}
\label{gmax_f2r_MS_GR}
f(r) = 1 -\frac{2 M}{r}-\frac{\kappa ^2 2^{-q} r^2}{3-4 q} \left(\frac{Q^2}{r^4}\right)^q.
\end{eqnarray}
It can be seen at a glance that $q\neq3/4$. Asymptotically,  $f(r)$ goes as
\[ \lim_{r\to\infty} f(r) =
\left \{
\begin{tabular}{ccc}
$1,$ & $q > 1/2,$\\
$1-\frac{1}{\sqrt{2}},$ & $q = 1/2,$\\
$-\infty,$ & $q < 1/2.$
\end{tabular}
\right \}
\]
Thus, the black hole is asymptotically flat only for $q>1/2$. The case of $q=1/2$ is unique as it goes to flat but differs from Minkowski at large distance~~\cite{Hassaine:2008pw, Hassaine:2007py}.
\begin{figure}
	\centering
	\includegraphics[height=8cm,keepaspectratio]{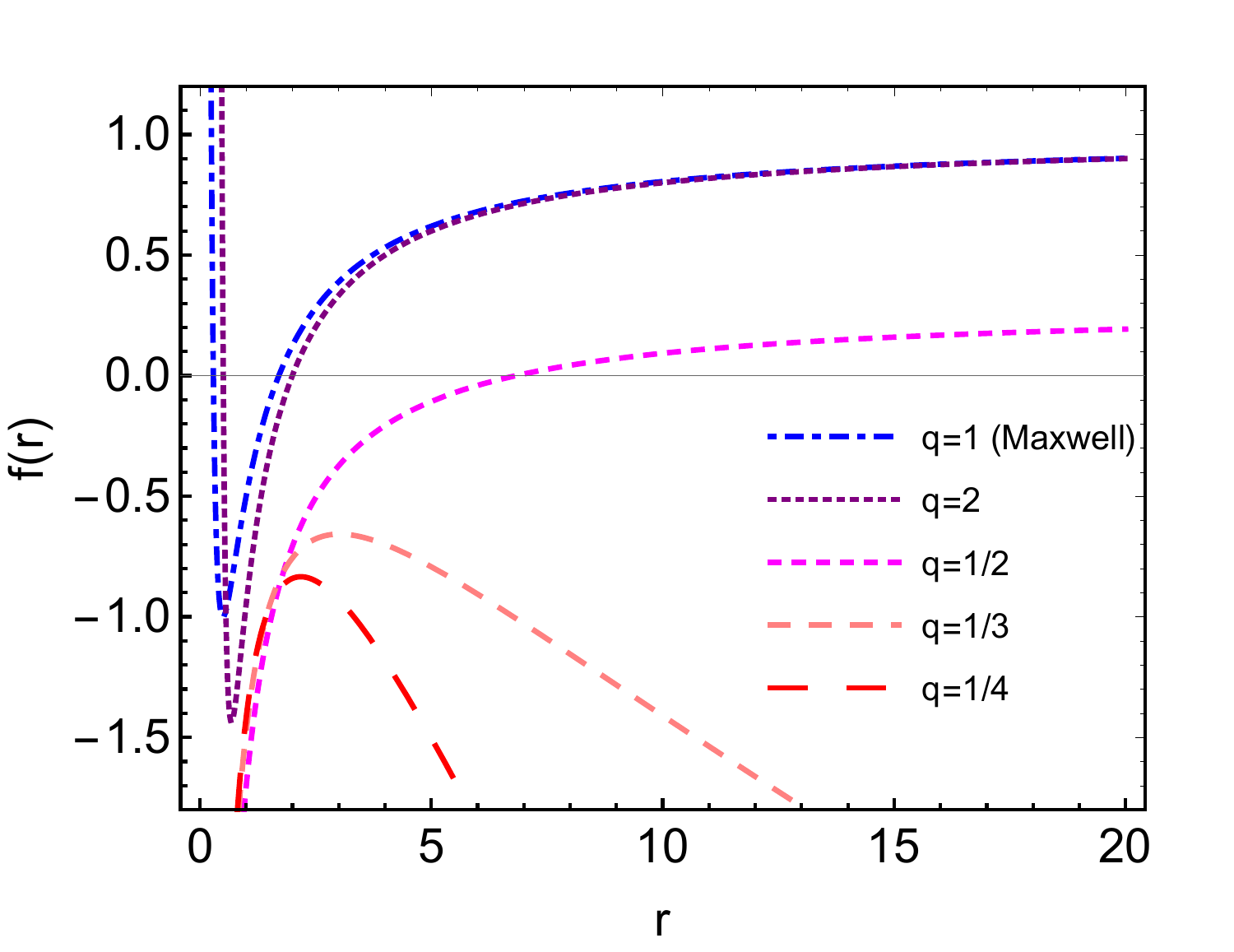} 
	\caption{Metric function $f(r)$ with $Q=M=1$.}
	\label{gr_max_ms_f}
\end{figure}

\subsection{Timelike Geodesics}

The corresponding $V_{eff}$ for massive test particles is
\begin{eqnarray}
\label{gmax_geo_MS_GR}
V_{eff}(r) = \frac{1}{2} \left(\frac{\mathbb{L}^2}{r^2}+1\right) \bigg[1 -\frac{2 M}{r}-\frac{\kappa ^2 2^{-q} r^2}{3-4 q} \left(\frac{Q^2}{r^4}\right)^q \bigg]-\frac{\mathbb{E}^2}{2}
\end{eqnarray}
Asymptotically,
\[ \lim_{r\to\infty} V_{eff}(r) =
\left \{
\begin{tabular}{ccc}
$0,$ & $q > 1/2,$\\
$-\frac{1}{2 \sqrt{2}},$ & $q = 1/2,$\\
$-\infty,$ & $q < 1/2.$
\end{tabular}
\right \}
\]
\begin{figure}[!ht]
	\centering
	\includegraphics[height=8cm,keepaspectratio]{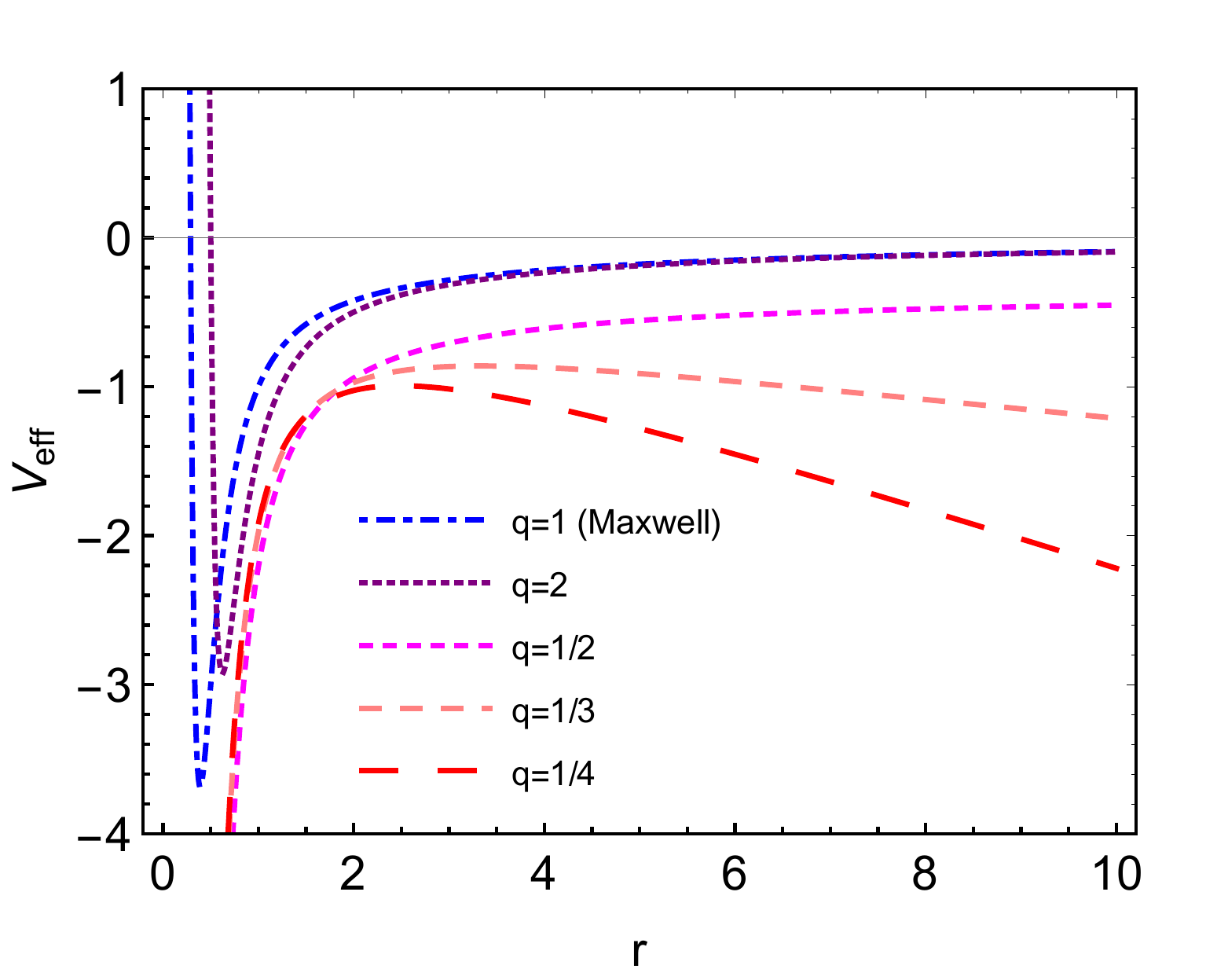} 
	\caption{The effective potential for massive particles with $M=1$, $Q=1$, and $\mathbb{E}=\mathbb{L}=1$.}
	\label{gr_max_ms_geo}
\end{figure}
The $V_{eff}$ behaves quite similarly with the metric, as shown in Fig.~\ref{gr_max_ms_geo}. No stable or unstable orbit exists outside the horizon.

\subsection{Null Geodesics}

The effective geometry in this model is given by
\begin{equation}
\label{gm_eff}
g^{\mu\nu}_{eff} = g^{\mu\nu} - \frac{4}{(q-1)\mathcal{F}} F^{\mu\alpha} F_{\alpha}^{\,\,\nu}.
\end{equation}
The line element can be written as Eq.~\eqref{gbi_eff} but with $h(r)\equiv(8 q-7)^{-1}$. The corresponding effective potential is
\begin{equation}
\label{gm_nullgeo}
V_{eff}(r) = \frac{ (8 q-7) \mathbb{L}^2}{2 r^2} \bigg( 1 -\frac{2 M}{r}-\frac{\kappa ^2 2^{-q} r^2}{3-4 q} \left(\frac{Q^2}{r^4}\right)^q \bigg) -\frac{\mathbb{E}^2}{2}.
\end{equation}
\begin{figure}[!ht]
	\centering
	\includegraphics[height=8cm,keepaspectratio]{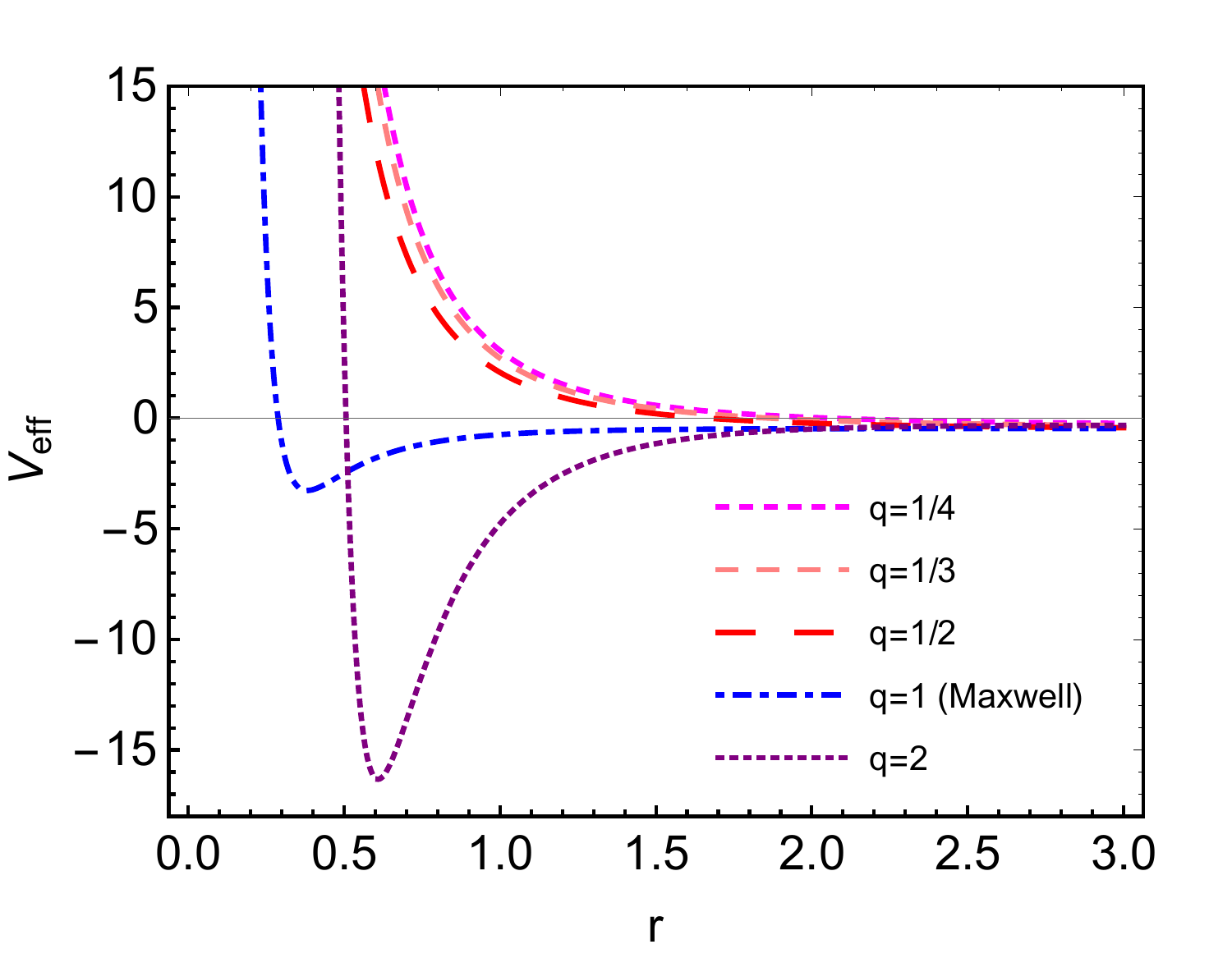} 
	\caption{The efective potential for ligt particles with $M=1$, $Q=1$, and $\mathbb{E}=\mathbb{L}=1$.}
	\label{gr_max_ms_null}
\end{figure}
In Fig.~\ref{gr_max_ms_null} we show the uninteresting result of $V_{eff}$, since no photon orbit (stable or unstable) exists either. In the following discussion we shall show that this power-law NLED model is problematic phenomenologically, at least in the weak-deflection limit.

\subsection{Deflection of Light}

For this model, it is easier to calculate the deflection angle through the first-order, rather than second-order as done previously, differential equation. Recalling the null geodesics equation \eqref{null_deff} and substituting $u=1/r$, the term $\dot{r}$ can be rewritten as
\begin{equation}
\label{rdot_deff}
\dot{r}^2 = \bigg( \frac{dr}{d\phi} \frac{d\phi}{d\tau}\bigg) = \mathbb{L}^2 \bigg( \frac{du}{d\phi} \bigg)^2.
\end{equation}
Eq.\eqref{null_deff} then becomes
\begin{equation}
\label{du2_deff}
\bigg( \frac{du}{d\phi} \bigg)^2 = \frac{1}{b^2} - \frac{f u^2}{h}.
\end{equation}
Defining $\sigma(u)\equiv b u (f/h)^{1/2}$, we obtain
\begin{equation}
\label{du_deff}
\frac{du}{d\phi} = \frac{1}{b} (1-\sigma^2)^{1/2}.
\end{equation}

As an example, let us take $q=2$. In the weak-field limit we obtain
\begin{equation}
\label{bdu_deff}
b du = \bigg( \frac{1}{3} + \frac{2 M \sigma }{9 b}-\frac{7 \kappa ^2 Q^4 \sigma ^6}{87480 b^6} \bigg) d\sigma.
\end{equation}
The total change of angle $\phi$ w.r.t. $u$ from infinity $(u=0)$ to the minimun radius $(u=u_0=1/b)$ and back to infinity can be written as
\begin{eqnarray}
\label{dphi_deff}
\delta \phi &=& 2 \int_{0}^{u_0} \frac{d\phi}{du} du = 2 \int_{0}^{u_0} b (1-\sigma^2)^{-1/2} du \nonumber \\
&=& 2 \int_{0}^{1} (1-\sigma^2)^{-1/2} \bigg(\frac{1}{3} + \frac{2 M \sigma }{9 b}-\frac{7 \kappa ^2 Q^4 \sigma ^6}{87480 b^6} \bigg) d\sigma.
\end{eqnarray}
The total deflection angle is defined as  $\Delta \phi\equiv\delta \phi - \pi$. Using this method we calculate the angles for several $q$ as follows:
\begin{eqnarray}
\label{gmax_def_MS_GR}
\Delta \phi(q=1) &=& \frac{4 M}{b} + \frac{3 \pi  \kappa  Q^2}{8 b^2}, \nonumber\\
\Delta \phi(q=2) &=& \frac{4 M}{9 b} + \pi  \left(-\frac{7 \kappa ^2 Q^4}{279936 b^6}-\frac{2}{3}\right), \nonumber\\
\Delta \phi(q=3) &=& \frac{4 M}{17 b}+\pi  \left(-\frac{77 \kappa ^2 Q^6}{5815734272 \sqrt{17} b}+\frac{1}{\sqrt{17}}-1\right).
\end{eqnarray}
\begin{figure}[!ht]
	\centering
	\includegraphics[height=8cm,keepaspectratio]{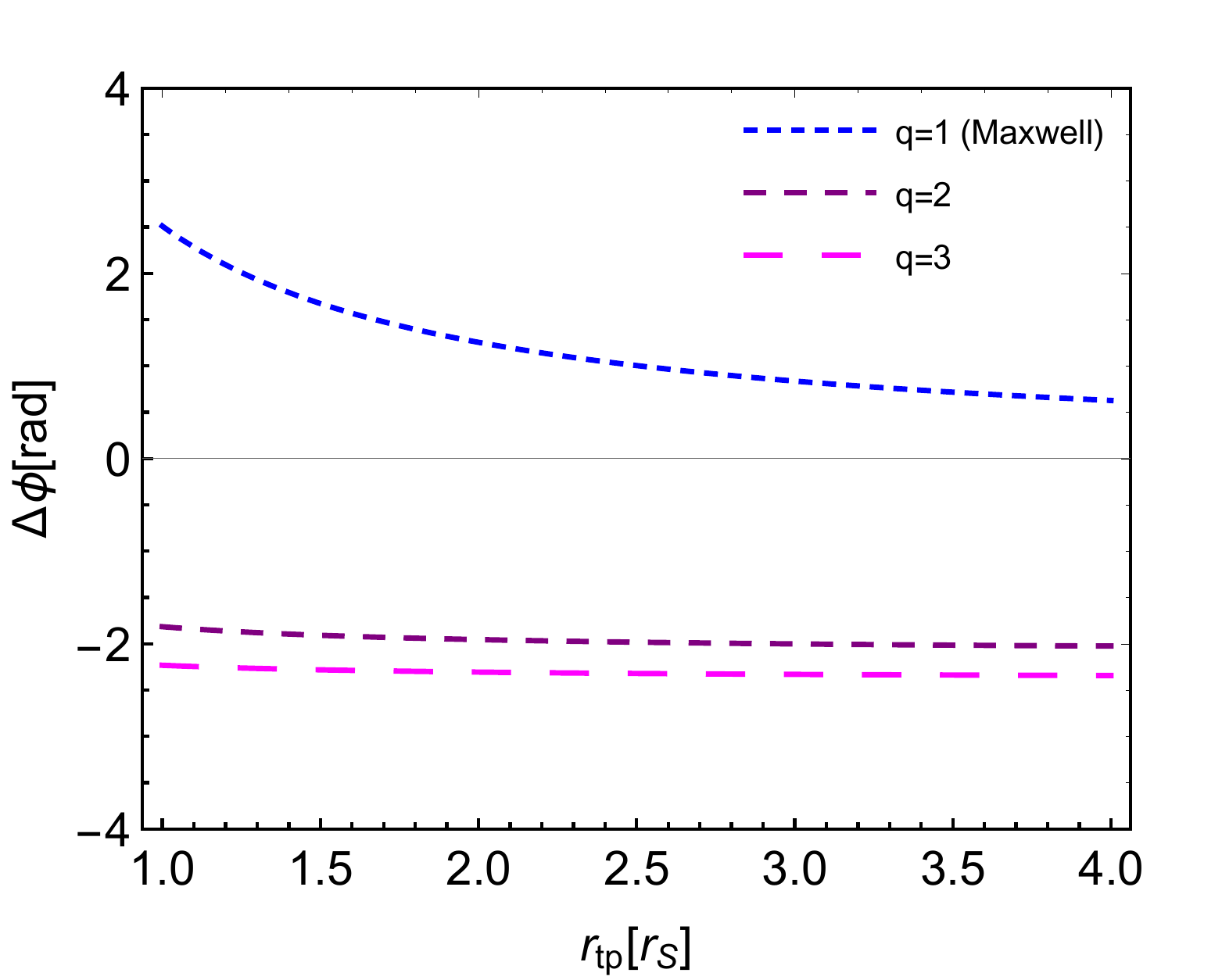} 
	\caption{Deflection angles for various number of $q$. Here we set $M=10 \,\, M_{\astrosun}$ and $Q=0.3$.}
	\label{gr_max_ms_def}
\end{figure}

As can be seen from the result above, unless $q=1$ the deflection angle does not reduce to Schwarszchild even in the limit of $Q\rightarrow0$. What is more problematic is that as shown i n Fig.~\ref{gr_max_ms_def} the angles are generically negative, and does not go to zero at large $r_{tp}$! This correspondence violation poses a doubt whether this result is physical or not. At best we can say that the weak-field approximation seems to break down for this model.

\section{Ayon-Beato-Garcia Black Hole}
\label{ABG_GR}

\begin{figure}
	\centering
	\begin{tabular}{c}
		\includegraphics[height=7cm,keepaspectratio]{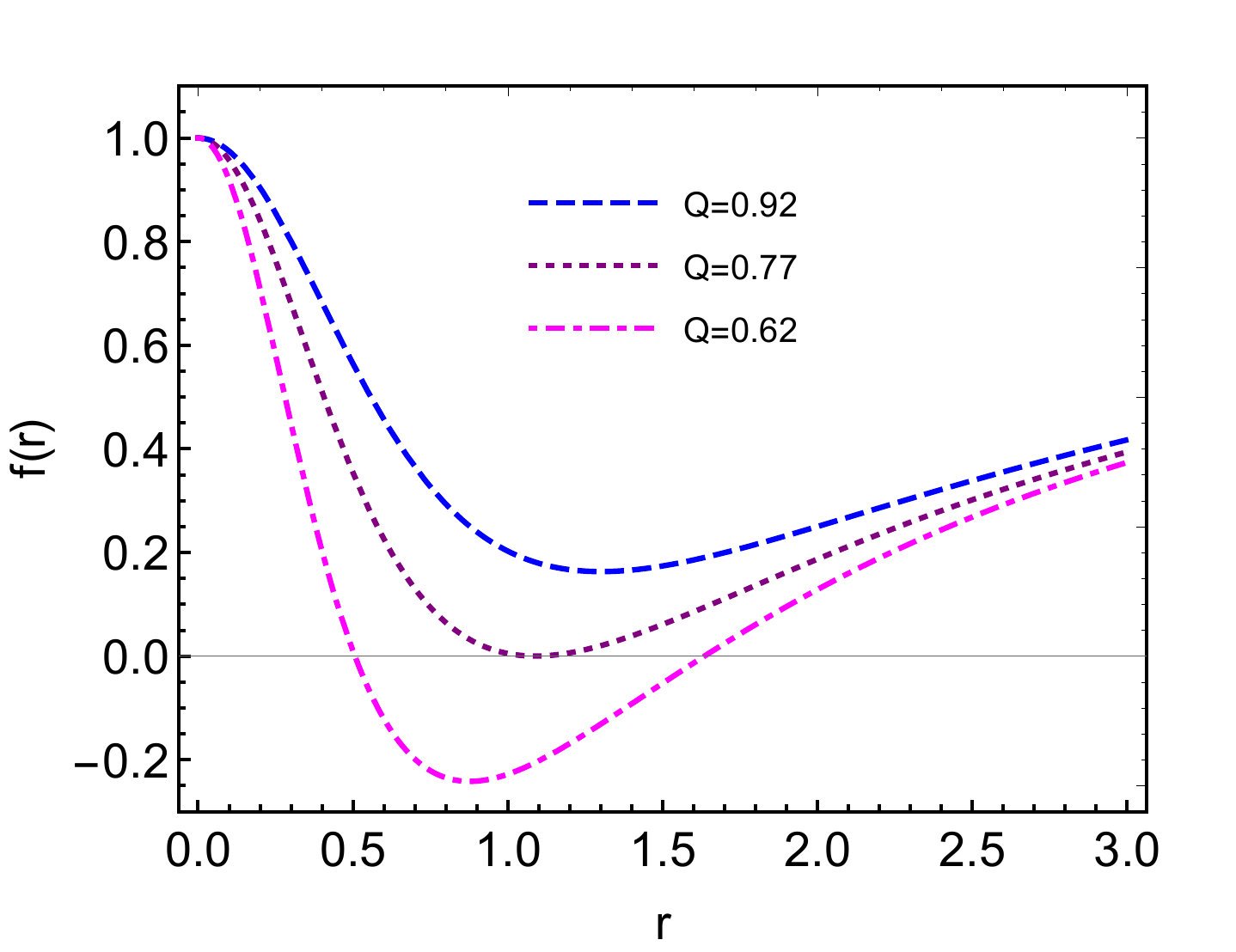}
	\end{tabular}
	\caption{Metric function for ABG black hole solution \eqref{abg_sol} with $M=1$ and several values of $Q$.}
	\label{gr_abg_ms_fr}
\end{figure}

In 1968 Bardeen~\cite{bardeen} in his seminal proceeding paper published his famous {\it regular black hole} solution,
\begin{equation}
\label{abg_sol}
f(r)=1-\frac{2 M r^2}{(r^2 + Q^2)^{3/2}}. 
\end{equation}
The metric is regular at the origin, as in Fig.~\ref{gr_abg_ms_fr}. The black hole regularity is ensured by the fact that the corresponding invariants are also regular everywhere. Ayon-Beato and Garcia~\cite{AyonBeato:2000zs} were the first to realize that such a solution can be interpreted as a black hole charged with NLED magnetic monopole whose Lagrangian is given by
\begin{equation}
\label{abg_L}
\mathcal{L} = -\frac{3}{s \kappa^2 Q^2} \bigg( \frac{\sqrt{2 Q^2\mathcal{F}}}{1 + \sqrt{2 Q^2\mathcal{F}}}\bigg)^{5/2},
\end{equation}
$s\equiv Q/2M$. Its electrically-charged counterpart solution was later proposed by Rodrigues and Silva~\cite{Rodrigues:2018bdc}. Being an NLED BH, we shall study the null geodesic structure of Ayon-Beato-Garcia (ABG) metric and investigate the stable photon orbits.

\subsection{Timelike and Null Geodesics}

\begin{figure}
	\centering
	\begin{tabular}{cc}
		\includegraphics[height=6.5cm,keepaspectratio]{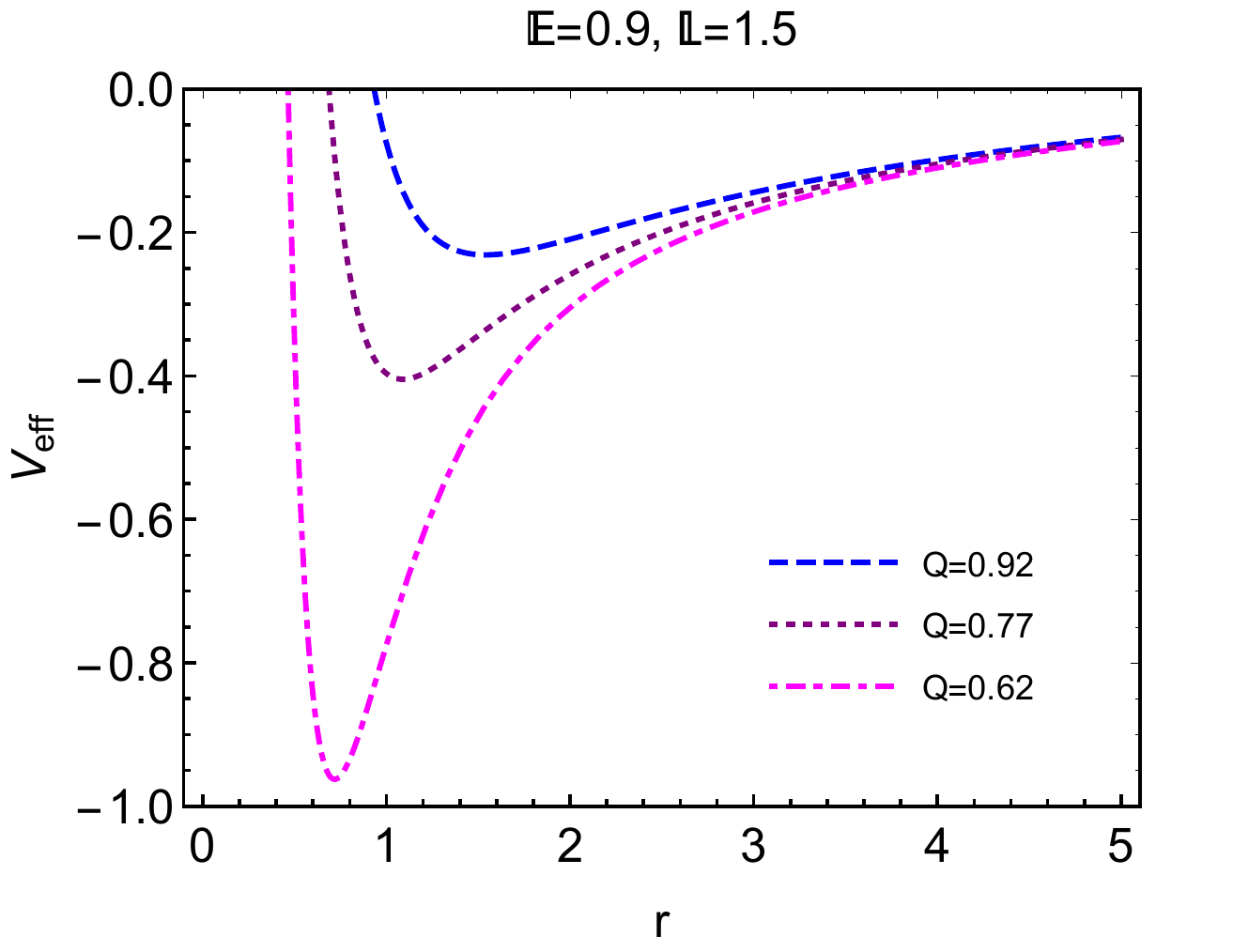} &
		\includegraphics[height=6.5cm,keepaspectratio]{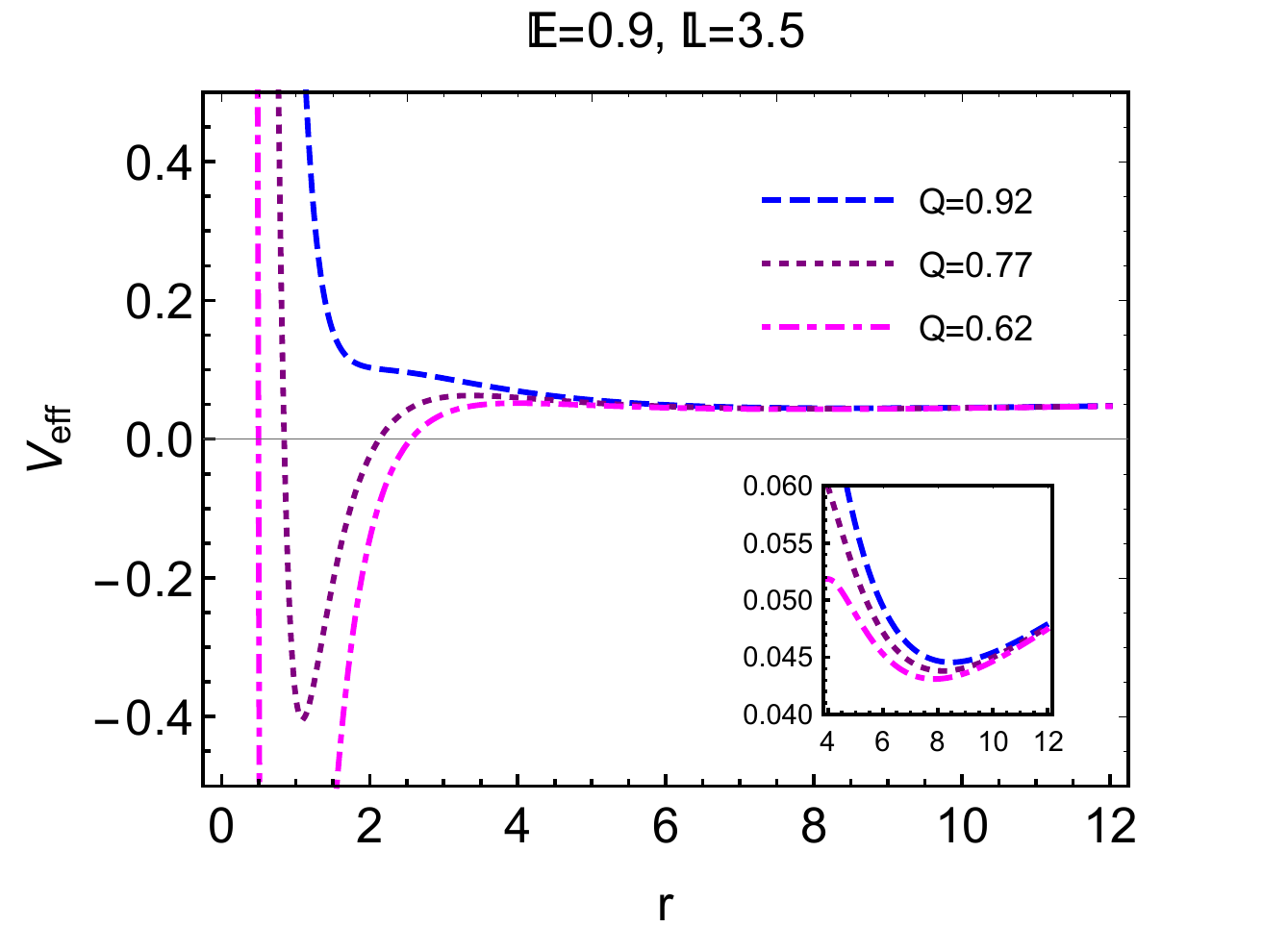}
	\end{tabular}
	\caption{The effective potential for massive particles with $M=1$ for various number of $Q$.}
	\label{gr_abg_ms_geo}
\end{figure}

The explicit form of the ABG's timelike geodesics can be written as
\begin{equation}
\label{abg_veff}
V_{eff} (r)= \frac{1}{2} \bigg(\frac{\mathbb{L}^2}{r^2} + 1\bigg) \bigg( 1-\frac{2 M r^2}{(r^2 + Q^2)^{3/2}} \bigg) -\frac{\mathbb{E}^2}{2}.
\end{equation}
This $V_{eff}$ allows (marginally) stable orbits for massive test particles, as can be seen in the right panel of Fig.~\ref{gr_abg_ms_geo}. 

While null geodesics structure of the original Bardeen spacetime was investigated by numerous authors (see, for example, \cite{Eiroa:2010wm, Ghaffarnejad:2014zva, Zhou:2011aa, Stuchlik:2014qja, Schee:2015nua}), none  assumes the NLED perspective. They thus neglected the photon's effective geometry and considered photon and graviton to follow the same null rays. Consequently no stable photon sphere is observed by an observer outside the horizon. Here we follow the ABG's perspective and found something novel.

\begin{figure}
	\centering
	\begin{tabular}{c}
		\includegraphics[height=8cm,keepaspectratio]{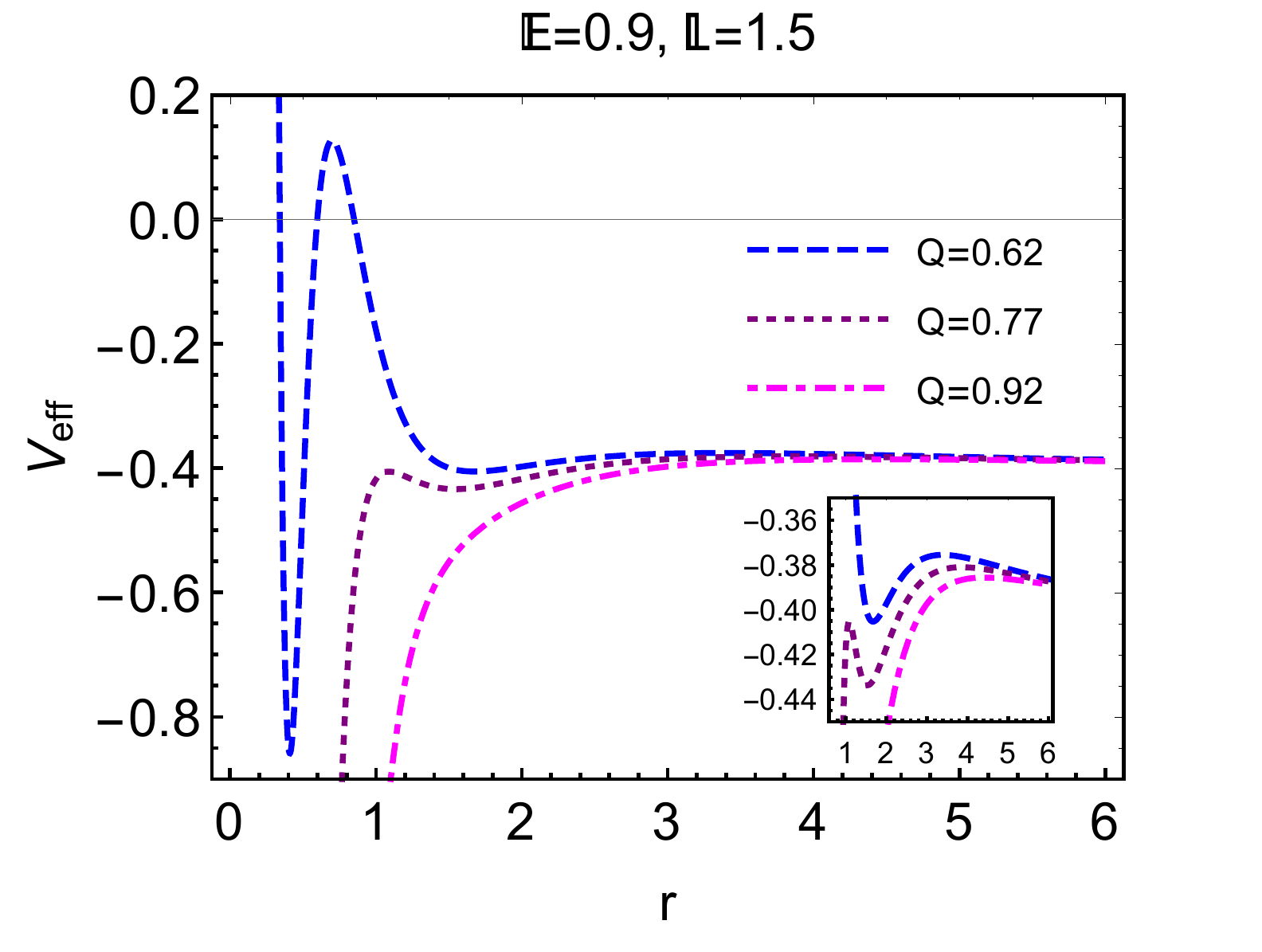}
	\end{tabular}
	\caption{The effective potential for light particles with $M=1$ for various number of $Q$. The corresponding equation is \eqref{abg_V_effg}}
	\label{gr_abg_ms_geo_eff}
\end{figure}

From Lagrangian~\ref{abg_L} the effective metric can be written as
\begin{equation}
\label{abg_eff}
g^{\mu\nu}_{eff} = g^{\mu\nu} + \frac{2 r^4 \left(6 Q^4+5 Q^2 r^2-1\right)}{Q^2 \left(Q^2+r^2\right)^2} F^{\mu\alpha} F_{\alpha}^{\nu}.
\end{equation}
Defining $h(r)\equiv\bigg( 1-\frac{2 \left(6 Q^4+5 Q^2 r^2-1\right)}{\left(Q^2+r^2\right)^2} \bigg)^{-1}$, the effective potential can be obtained,
\begin{equation}
\label{abg_V_effg}
V_{eff}(r) = \frac{\mathbb{L}^2}{2 r^2} \bigg( 1-\frac{2 \left(6 Q^4+5 Q^2 r^2-1\right)}{\left(Q^2+r^2\right)^2} \bigg) \bigg( 1-\frac{2 M r^2}{(r^2 + Q^2)^{3/2}} \bigg) -\frac{\mathbb{E}^2}{2}.
\end{equation}
The circular orbit radii satisfies
\begin{eqnarray}
0&=&\frac{\mathbb{L} \left(\frac{4 r \left(12 Q^4+10 Q^2
   r^2-2\right)}{\left(Q^2+r^2\right)^3}-\frac{20 Q^2 r}{\left(Q^2+r^2\right)^2}\right)
   \left(1-\frac{2 M r^2}{\left(Q^2+r^2\right)^{3/2}}\right)}{2 r^2}\nonumber\\
   &&+\frac{\mathbb{L} \left(1-\frac{12 Q^4+10 Q^2 r^2-2}{\left(Q^2+r^2\right)^2}\right) \left(\frac{6 M
   r^3}{\left(Q^2+r^2\right)^{5/2}}-\frac{4 M r}{\left(Q^2+r^2\right)^{3/2}}\right)}{2
   r^2}\nonumber\\ 
   &&-\frac{\mathbb{L} \left(1-\frac{12 Q^4+10 Q^2 r^2-2}{\left(Q^2+r^2\right)^2}\right) \left(1-\frac{2
   M r^2}{\left(Q^2+r^2\right)^{3/2}}\right)}{r^3}.
  \end{eqnarray}
Finding its analytical roots is not illuminating. We thus determine the $r_{SCO}$ by studying Fig.~\ref{gr_abg_ms_geo_eff}. It can be seen that for the case of extremal and two horizons there are minima, albeit of little depth, whose $r_{SCO}>r_h$. This means that ABG BH allows stable photon orbits. The detail numerical values of the corresponding radii is shown in Table.~\ref{abg_rEH_null}. These stability is considered to be marginally stable, since the depth of minima is quite shallow. Nevertheless, it is not a problem classically.
\begin{table}
	\setlength{\tabcolsep}{1em}
	\begin{tabular}{||c | c | c | c||} 
		\hline
		$Q$ & $r_{h}$ & $r_{UCO}$ & $r_{SCO}$ \\ [0.5ex] 
		\hline\hline
		0.62 & 0.5065 and 1.6349 &  0.6994 and 3.3887 & 0.4097 and 1.6743 \\ 
		\hline
		0.77 & 1.0887 & 1.0897 and 3.8556 & 1.54353 \\
		\hline
		0.92 & - & 4.4158 & - \\
		\hline
	\end{tabular}
	\caption{Comparation of the radius of event horizon of the extremal case ($r_{EH}$) $Q$, next to its corresponding values of $r_{SCO}$ and $r_{SCO}$ of the null geodesics.}
	\label{abg_rEH_null}
\end{table}

\subsection{Deflection of Light}

\begin{figure}[!ht]
	\centering
	\includegraphics[height=9cm,keepaspectratio]{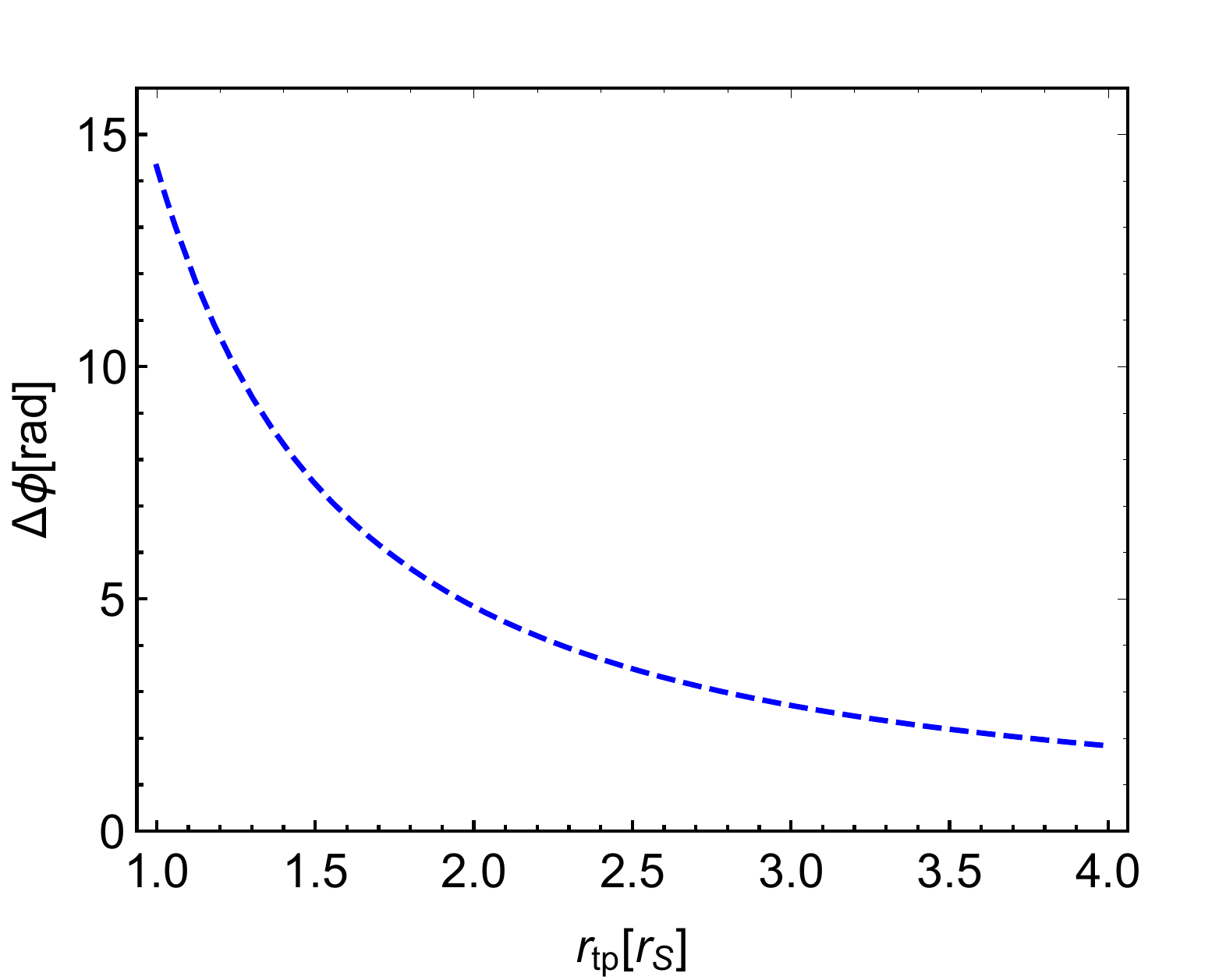} 
	\caption{Deflection angles of ABG black holes.}
	\label{gr_abg_ms_def}
\end{figure}

Finally, let us calculate the deflection angle of photon off the ABG black hole which, to the best of our knowledge, has not been studied in the literature. In the limit of small $Q$ the inverse radial distance is, approximately,
\begin{eqnarray}
\label{abg_u}
u&\simeq& \frac{u_0^3}{32 \alpha ^2} \big(12 \sqrt{\alpha } \phi  \sin \left(\sqrt{\alpha } \phi \right) \left((\alpha -1) \alpha  b_0^2+10 M^2\right) \nonumber \\
&& +\cos \left(\sqrt{\alpha } \phi \right) \left(7 (\alpha -1) \alpha  b_0^2+54 M^2\right)+\cos \left(3 \sqrt{\alpha } \phi \right) \left(6 M^2-(\alpha -1) \alpha  b_0^2\right)\big) \nonumber \\
&& +\frac{M u_0^2 }{4 \alpha } \left(9 \sin ^2\left(\sqrt{\alpha } \phi \right)+\sin \left(\sqrt{\alpha } \phi \right) \sin \left(3 \sqrt{\alpha } \phi \right)+4 \cos ^4\left(\sqrt{\alpha } \phi \right)\right) +u_0 \cos \left(\sqrt{\alpha } \phi \right).\nonumber\\
\end{eqnarray}
where we define $\alpha\equiv1-\frac{20 Q^2}{b_0^2}$. For $\alpha \approx 1$ the weak deflection angle is given by\begin{equation}
\Delta \phi_{weak}\approx\frac{8 M}{r_{tp}} + \frac{15 \pi  M^2}{2 r_{tp}^2}+\mathcal{O}\left(Q^2\right).
\end{equation}
The zeroth-order is twice the Schwarzschild's angle. This is because ABG describes a regular (Bardeen) black hole which is distinct from Schwarzschild. The dependence of $\Delta\phi$ on $r_{tp}$ is shown in Fig.~\ref{gr_abg_ms_def}. This might be used to distinguish ABG/Bardeen's signature from other ordinary black hole. 

\section{Conclusion}
\label{conclude}

This work is intended to investigate null geodesic of several NLED black holes and its phenomenological aspects. In this work we specifically consider three polynomial-type NLED models: the generalized BI model (we dubbed it the Kruglov-BI model), the power-law model, and the Ayon-Beato-Garcia model used in Bardeen black hole. Each has been extensively studied by a large number of authors. What is left uninvestigated, and this is our main result, is the behaviour of photon around them. Due to the field discontinuity, photon follows the null geodesic that is different from graviton in NLED theories. Our simple investigation shows novel results.

In the first model, we show that in the extremal limit of some weak-coupling $\beta$ the black hole allows stable physical photon orbits. By {\it physical} we mean that the orbits lie outside the horizon. For the strong-coupling $\beta<1$, generically for $q<1$ the black hole only possesses one horizon, and we show that there is a range of parameters that also allows this case to have stable physical orbits for photon. The results are genuine, since extremal RN has stable photon orbit with $r_{SCO}$ coincides with $r_{EH}$. It is not clear how stable this is since any small perturbation might collapse the photon inside the horizon. Now, the possibility of having $r_{SCO}>r_{EH}$ in NLED case evades such concern. Not only we now have stable circular orbits, but also there exists a family of bounded orbits parametrized by two radii, $r_+$ and $r_-$, as long as $r_-\geq r_{EH}$. In Newtonian gravity these closed orbits would correspond to ellipse. In GR, however, to determine the orbits we must explicitly solve the null geodesic equation. This is left for our future investigation. 

For the second model, we found problematic phenomenological results since the weak deflection angle does not coincide with Schwarzschild even in the chargeless limit ($Q\rightarrow0$). We argue that such model is unphysical, even though we do not eliminate the possibility that our weak-gravitational field approach might be breakdown for this model. At best we can say that the weak-field analysis fails and one must resort to the full strong deflection analysis to get satisfactory answer.

The last model deals with the ABG metric that gives rise to regular black hole. Regular black hole was first proposed, to our best knowledge, by Bardeen. Later Ayon-Beato and Garcia realized that such solution can be perceived as black hole endowed with NLED magnetic charge. An interesting aspect is that, while the metric solution and the timelike geodesic between Bardeen and ABBG are equivalent, the null geodesic is not. Perceived as original Bardeen, regular BH has the usual property that photon follows its null geodesic. Study of its null geodesic showed that the corresponding $V_{eff}$ possesses singularity~\cite{Novello:2000km}; therefore it is futile to talk about photon orbit around Bardeen BH. But looking from the NLED perspective, the matter becomes non-trivial. A test photon follows its effective geometry, and our investigation reveals that it is non-singular. Similar to the Kruglov-BI case the ABG black hole also allows photon sphere and other bounded orbits outside the horizon. Our calculation on the weak deflection angle shows that up to zeroth-order it is twice the Schwarzschild value. While we argue that this might be caused by the fact that the two have different nature regarding the singularity (thus, the effective null geodesic outside the horizon is influenced by the nature inside it), we also realize that the strong deflection formalism is needed to have a conclusive hypothesis. 


\acknowledgments
We thank Byon Jayawiguna, Reyhan Lambaga, and Ilham Prasetyo for fruitful discussions. We are also grateful for the referee's comments and suggestions to the first manuscript. This work is partially funded by Hibah PUTI Q1 UI 2020.


\end{document}